\documentclass[10pt,journal,compsoc]{IEEEtran}
\usepackage{algpseudocode}

\usepackage{graphicx}
\usepackage{float}
\usepackage{array}
\usepackage{tabularx}
\usepackage{balance}
\usepackage{marvosym}
\usepackage{amssymb}
\usepackage{amsmath}
\usepackage{multirow}
\usepackage{array}
\usepackage{enumitem}
\usepackage{booktabs} 
\usepackage[ruled,vlined]{algorithm2e}
\usepackage{amsfonts}
\usepackage{graphicx}
\usepackage{subfig}
\usepackage{textcomp}
\usepackage[dvipsnames, svgnames, x11names]{xcolor} 
\usepackage{pifont}
\usepackage{url}
\usepackage{bm}
\usepackage{makecell}
\usepackage{adjustbox}
\usepackage{caption}

\makeatletter

\DeclareMathOperator{\sign}{sign}

\begin{document}
	
	\title{XSimGCL: Towards Extremely Simple Graph Contrastive Learning for Recommendation}
	
	\author{Junliang Yu, Xin Xia, Tong Chen, Lizhen Cui, Nguyen Quoc Viet Hung, Hongzhi~Yin$^{*}$
		\IEEEcompsocitemizethanks{
			\IEEEcompsocthanksitem J. Yu, X. Xia, T. Chen, and H. Yin are with the School of Information Technology and Electrical Engineering, The University of Queensland, Brisbane, Queensland, Australia.\protect\\
			E-mail: \{jl.yu, x.xia, tong.chen, h.yin1\}@uq.edu.au
			\IEEEcompsocthanksitem H. Nguyen is with the Institute for Integrated and Intelligent Systems, Griffith University, Gold Coast, Australia.\protect\\
			E-mail: quocviethung1@gmail.com
			\IEEEcompsocthanksitem Lizhen Cui is with the School of Software, Shandong University, Jinan, China.\protect\\
			E-mail: clz@sdu.edu.cn						
		}
		\thanks{$^{*}$Corresponding author.}
	}

	\markboth{IEEE TRANSACTIONS ON KNOWLEDGE AND DATA ENGINEERING}%
	{Shell \MakeLowercase{\textit{et al.}}: Bare Demo of IEEEtran.cls for Computer Society Journals}

	\IEEEtitleabstractindextext{%
		\begin{abstract} 
			Contrastive learning (CL) has recently been demonstrated critical in improving recommendation performance. The underlying principle of CL-based recommendation models is to ensure the consistency between representations derived from different graph augmentations of the user-item bipartite graph. This self-supervised approach allows for the extraction of general features from raw data, thereby mitigating the issue of data sparsity. Despite the effectiveness of this paradigm, the factors contributing to its performance gains have yet to be fully understood. This paper provides novel insights into the impact of CL on recommendation. Our findings indicate that CL enables the model to learn more evenly distributed user and item representations, which alleviates the prevalent popularity bias and promoting long-tail items. Our analysis also suggests that the graph augmentations, previously considered essential, are relatively unreliable and of limited significance in CL-based recommendation. Based on these findings, we put forward an e\textbf{X}tremely \textbf{Sim}ple \textbf{G}raph \textbf{C}ontrastive \textbf{L}earning method (\textbf{XSimGCL}) for recommendation, which discards the ineffective graph augmentations and instead employs a simple yet effective noise-based embedding augmentation to generate views for CL. A comprehensive experimental study on four large and highly sparse benchmark datasets demonstrates that, though the proposed method is extremely simple, it can smoothly adjust the uniformity of learned representations and outperforms its graph augmentation-based counterparts by a large margin in both recommendation accuracy and training efficiency. The code and used datasets are released at \url{https://github.com/Coder-Yu/SELFRec}.
		\end{abstract}
		
		}

	\maketitle

	\IEEEdisplaynontitleabstractindextext

	%
	\IEEEpeerreviewmaketitle	
	
\begin{figure}[t]
		\centering
		\includegraphics[width=.45\textwidth]{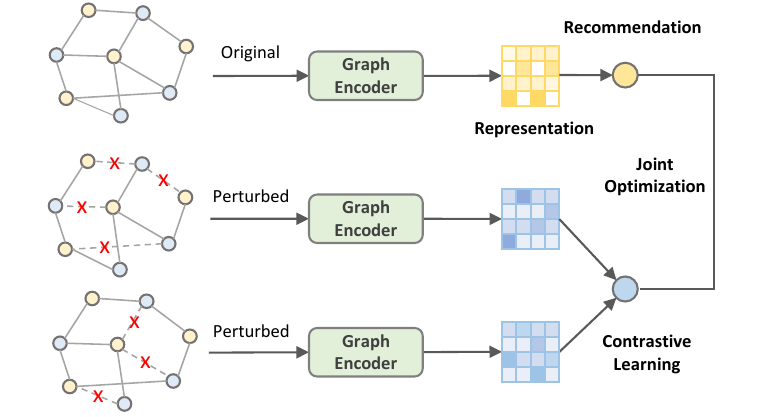}
		\caption{Graph contrastive learning with edge dropout for recommendation.}
		\label{fig:gcl}
		\vspace{-10pt}				
	\end{figure}

	\IEEEraisesectionheading{\section{Introduction}\label{sec:introduction}}
	\IEEEPARstart{T}{h}e recent resurgence of Contrastive Learning (CL) \cite{jaiswal2021survey,liu2020self,yu2022survey} in various domains of deep learning has led to a series of breakthroughs \cite{you2020graph,chen2020simple,gao2021simcse,he2020momentum,grill2020bootstrap}. Since the ability of CL to learn general features from unlabeled raw data has proven to be an effective solution for the issue of data sparsity \cite{singh2020scalability,TamWTYH17,chen2020sequence}, it has also sparked significant advancements in the field of recommendation. A surge of enthusiasm on CL-based recommendation \cite{wujc2021self,yu2021self,xia2020self,zhou2020s,zhou2021contrastive,yu2021socially,lin2022improving} has recently been witnessed, followed by a string of promising outcomes. The paradigm of CL-based recommendation can be defined as a two-step process: firstly augmenting the original user-item bipartite graph with structural perturbations (e.g. edge or node dropout at a specific rate), and secondly maximizing the consistency of representations learned from different graph augmentations under a joint learning framework \cite{yu2022survey} (shown in Fig. \ref{fig:gcl}). \par

	Despite the demonstrated effectiveness of this paradigm, the underlying mechanism driving performance gains remains elusive. Intuitively, encouraging the agreement between related graph augmentations can lead to the learned representations invariant to slight structural perturbations and capture the essential information of the original user-item bipartite graph \cite{jaiswal2021survey,bachman2019learning}. However, several recent studies have reported unexpected results, indicating that the performance of CL-based recommendation models is not sensitive to the edge dropout rates of graph augmentations, and that even high dropout rates (e.g. 0.9) can still improve model performance \cite{yu2021socially,zhou2021selfcf,DBLP:conf/sigir/LeeKJPY21}. This naturally raises an intriguing and bedrock question: \textit{Are graph augmentations really a necessity for CL-based recommendation models?} \par

	To answer this question, we initially performed experiments both with and without graph augmentations, and evaluated their respective performances. Our findings indicate that while there is a minor decline in performance when graph augmentations are detached, the real decisive factor lies in the representation learning. Upon visual inspection of these representations, it was determined that the contrastive loss InfoNCE \cite{oord2018representation}, is the primary contributor to improved performance. Optimizing it leads to a more even distribution of user/item representations, mitigating the impact of popularity bias \cite{chen2020bias} and promoting long-tail items. On the other hand, though not as effective as expected, some types of graph augmentations indeed improve the recommendation performance. However, a lengthy trial-and-error is needed to identify the most effective ones. Otherwise a random selection may degrade the recommendation performance. Besides, it should be aware that repeatedly creating graph augmentations and constructing adjacency matrices bring extra expense to model training. Considering these limitations, it may be more practical to pursue alternative augmentations that are both more effective and efficient. A follow-up question then arises: \textit{Are there any more effective and efficient augmentation approaches?}
	\par
	In our previous study \cite{yu2022graph}, we had given an affirmative response to this question. Building upon our conclusion that learning more evenly distributed representations is critical for enhancing recommendation performance, we proposed a graph-augmentation-free CL method which makes the uniformity more controllable, and named it \textbf{SimGCL} (short for \textbf{Sim}ple \textbf{G}raph \textbf{C}ontrastive \textbf{L}earning). SimGCL conforms to the paradigm presented in Fig. \ref{fig:gcl}, but eliminates ineffective graph augmentations and instead implements a more efficient representation-level data augmentation through the addition of uniform noises to the learned representations. Empirical results demonstrated that this noise-based augmentation can directly regularize the embedding space towards a more even representation distribution. Moreover, by controlling the magnitude of noises, SimGCL allows for the smooth adjustment of representation uniformity. Benefitting from these characteristics, SimGCL shows superiorities over its graph augmentation-based counterparts in both recommendation accuracy and training efficiency.
	\par
\begin{figure}[t]
		\centering
		\captionsetup{justification=centering}
		\includegraphics[width=.48\textwidth]{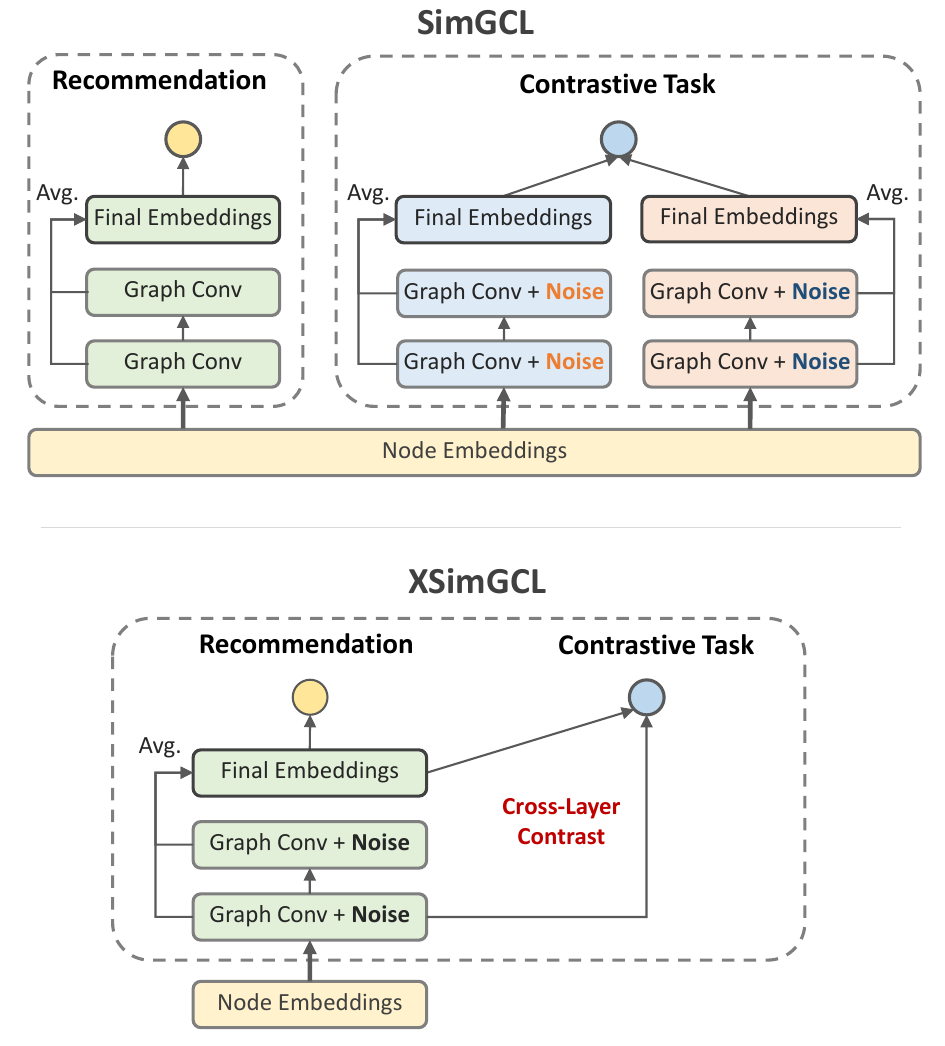}
		\caption{The architectures of SimGCL and XSimGCL. }
		\label{fig:architecture}				
	\end{figure}

	However, in spite of these advantages, \textbf{the cumbersome architecture of SimGCL renders it less than perfect}. In addition to the forward/backward pass for the recommendation task, it necessitates two additional forward and backward passes for the contrastive task within each mini-batch, as shown in Fig. \ref{fig:architecture}. Actually, this is a universal problem for all the CL-based recommendation models \cite{wujc2021self,xie2022contrastive,yu2021socially,zhang2021double,yao2021self} following the paradigm in Fig. \ref{fig:gcl}. What makes it worse is that these methods require all nodes in the user-item bipartite graph to be present during training, which increases the computational cost to nearly triple that of conventional recommendation models. This flaw greatly hinders the scalability of CL-based models.
	\par

	In order to address this issue, in this work we put forward an e\textbf{X}tremely \textbf{Sim}ple \textbf{G}raph \textbf{C}ontrastive \textbf{L}earning method (\textbf{XSimGCL}) for recommendation. XSimGCL builds upon SimGCL's noise-based augmentation approach while streamlining the computation process through a shared single pass that unifies the recommendation and contrastive tasks. The implementation of XSimGCL is depicted in Fig. \ref{fig:architecture}. To be specific, both SimGCL and XSimGCL are fed with the same input: the initial embeddings and the adjacency matrix. The difference is that SimGCL contrasts two final representations learned through different forward passes and relies on the ordinary representations for recommendation, whereas XSimGCL uses the same perturbed representations for both tasks, and replaces the final-layer contrast in SimGCL with the cross-layer contrast. This design makes XSimGCL nearly as lightweight as the conventional recommendation model LightGCN \cite{he2020lightgcn}. And to top it all, XSimGCL even outperforms SimGCL with its simpler architecture. 
	\par
	The current work extends the findings of our previous study \cite{yu2022graph} and presents the following contributions:
	\begin{itemize}[leftmargin=*]
		\item We elucidate the beneficial effects of contrastive learning (CL) on graph recommendation models, specifically by learning representations that are more uniformly distributed, where the InfoNCE loss holds greater significance than graph augmentations.
		\item We propose a simple yet effective noise-based augmentation approach, which enables the smooth adjustment of the uniformity of learned representations.
		\item We put forward a novel CL-based recommendation model XSimGCL that surpasses its predecessor SimGCL in terms of effectiveness and efficiency. Additionally, we provide theoretical analysis explaining the superiority of XSimGCL through the lens of graph spectrum.
		\item We conduct a comprehensive experimental study on four large and highly sparse benchmark datasets (three of which were not used in our preliminary study) to demonstrate that XSimGCL is an ideal alternative of its graph augmentation-based counterparts.	
	\end{itemize} 
	The rest of this paper is organized as follows. Section \ref{sec:revisit} investigates the necessity of graph augmentations in the contrastive recommendation and explores how CL enhances recommendation. Section \ref{sec:XSimGCL} proposes the noise-based augmentation approach and the CL-based recommendation model XSimGCL. The experimental study is presented in Section \ref{sec:experiments}. Section \ref{sec:related} provides a brief review of the related literature. Finally, we conclude this work in Section \ref{sec:conclusion}.

\section{Revisiting Graph CL for Recommendation}
\label{sec:revisit}
\subsection{Contrastive Recommendation with Graph Augmentations}
Generally, data augmentations are the prerequisite for CL-based recommendation models \cite{zhou2020s,wujc2021self,yu2021self,xie2022contrastive}. In this section, we investigate the widely used dropout-based augmentations on graphs \cite{wujc2021self,you2020graph}. It is assumed that the learned representations which are invariant to partial structure perturbations are high-quality. We target a representative state-of-the-art CL-based recommendation model SGL \cite{wujc2021self}, which performs the node/edge dropout to augment the user-item graph. The joint learning scheme in SGL is formulated as:
\begin{equation}
	\label{joint}
	\mathcal{L} = \mathcal{L}_{rec} + \lambda\mathcal{L}_{cl},
\end{equation}
which consists of the recommendation loss $\mathcal{L}_{rec}$ and the contrastive loss $\mathcal{L}_{cl}$. Since the goal of SGL is to recommend items, the CL task plays an auxiliary role and its effect is modulated by a hyperparameter $\lambda$. As for the instantiations of these two losses, the standard BPR loss \cite{rendle2009bpr} and the InfoNCE loss \cite{oord2018representation} are adopted in SGL for recommendation and CL, respectively. The standard BPR loss is defined as:
\begin{equation}
	\label{eq:bpr}
	\mathcal{L}_{rec} = -\sum_{(u,i) \in \mathcal{B}}\log\left(\sigma(\mathbf{e}_{u}^{\top}\mathbf{e}_{i}-\mathbf{e}_{u}^{\top}\mathbf{e}_{j})\right),
\end{equation}
where $\sigma$ is the sigmoid function, $\mathbf{e}_{u}$ is the user representation, $\mathbf{e}_{i}$ is the representation of an item that user $u$ has interacted with, $\mathbf{e}_{j}$ is the representation of a randomly sampled item, and $\mathcal{B}$ is a mini-batch. The InfoNCE loss \cite{oord2018representation} is formulated as:
\begin{equation}
	\label{loss:cl}
	\mathcal{L}_{cl}=\sum_{i \in \mathcal{B}}-\log \frac{\exp (\mathbf{z}_{i}^{\prime\top}\mathbf{z}_{i}^{\prime \prime} / \tau)}{\sum_{j \in \mathcal{B}} \exp (\mathbf{z}_{i}^{\prime\top}\mathbf{z}_{j}^{\prime \prime} / \tau)},
\end{equation}
where $i$ and $j$ are users/items in $\mathcal{B}$, $\mathbf{z}^{\prime}_{i}$ and $\mathbf{z}^{\prime \prime}_{i}$ are $L_{2}$ normalized representations learned from two different dropout-based graph augmentations (namely $\mathbf{z}_{i}^{\prime}=\frac{\mathbf{e}_{i}^{\prime}}{\|\mathbf{e}_{i}^{\prime}\|_{2}}$), and $\tau > 0$ (e.g., 0.2) is the temperature which controls the strength of penalties on hard negative samples. The InfoNCE loss encourages the consistency between $\mathbf{z}_{i}^{\prime}$ and $\mathbf{z}_{i}^{\prime \prime}$ which are the positive sample of each other, whilst minimizing the agreement between $\mathbf{z}_{i}^{\prime}$ and $\mathbf{z}_{j}^{\prime \prime}$, which are the negative samples of each other. Optimizing the InfoNCE loss is actually maximizing a tight lower bound of mutual information.
\par
To learn representations from the user-item graph, SGL employs LightGCN \cite{he2020lightgcn} as its encoder, whose message passing process is defined as:
\begin{equation}
	\label{eq:lightgcn}
	\mathbf{E}=\frac{1}{1+L}(\mathbf{E}^{(0)}+\bar{\mathbf{A}}\mathbf{E}^{(0)}+...+\bar{\mathbf{A}}^{L}\mathbf{E}^{(0)}),
\end{equation} 
where $\mathbf{E}^{(0)}\in\mathbb{R}^{|N|\times d}$ is the node embeddings to be learned, $\mathbf{E}$ is the initial final representations for prediction, $|N|$ is the number of nodes, $L$ is the number of layers, and $\bar{\mathbf{A}}\in\mathbb{R}^{|N|\times |N|}$ is the normalized undirected adjacency matrix without self-connection. By replacing $\bar{\mathbf{A}}$ with the adjacency matrix of the corrupted graph augmentations $\tilde{\mathbf{A}}$, $\mathbf{z}^{\prime}$ and $\mathbf{z}^{\prime \prime}$ can be learned via Eq. (\ref{eq:lightgcn}). It should be noted that in every epoch, $\tilde{\mathbf{A}}$ is reconstructed. For the sake of brevity, here we just present the core ingredients of SGL and LightGCN. More details can be found in the original papers \cite{wujc2021self,he2020lightgcn}.

\begin{figure*}[t]	
	\centering
	\subfloat[Distribution learned from Yelp2018]{%
	  \includegraphics[clip,width=\textwidth]{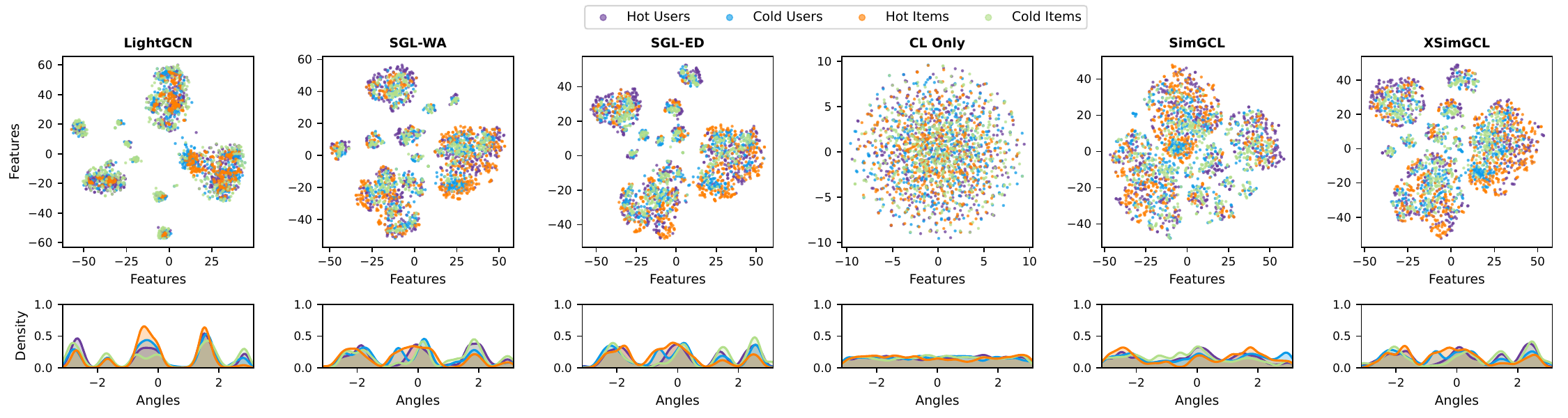}%
	}	
	\\
	\subfloat[Distribution learned from Amazon-Kindle]{%
	\includegraphics[clip,width=\textwidth]{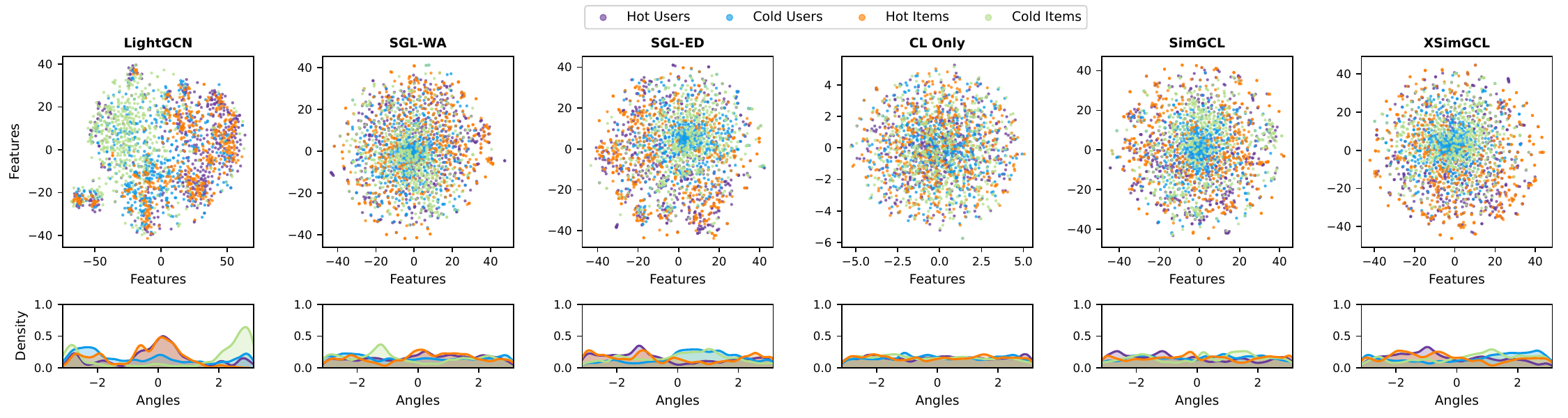}%
  }	
  \\
	\subfloat[Distribution learned from Alibaba-iFashion]{%
	  \includegraphics[clip,width=\textwidth]{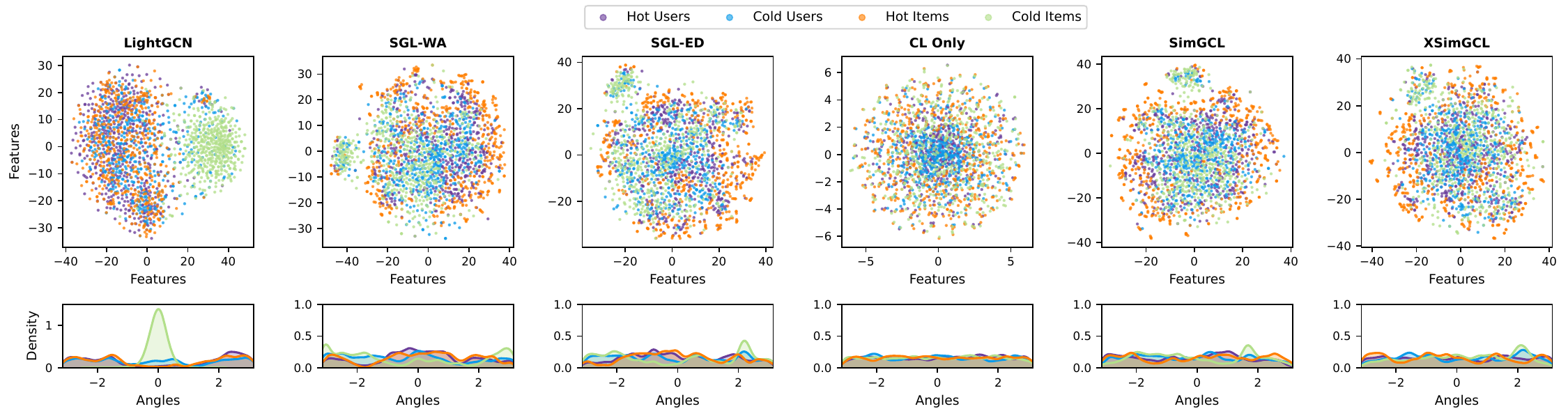}%
	}	
	\caption{The distribution of representations learned from three datasets. The top of each figure plots the learned 2D features and the bottom of each figure plots the Gaussian kernel density estimation of atan2(y, x) for each point (x,y) $\in \mathcal{S}^{1}$}
	\label{figure:dist} 
\end{figure*}

\subsection{Necessity of Graph Augmentations}
The findings reported in recent studies \cite{yu2021socially,zhou2021selfcf,DBLP:conf/sigir/LeeKJPY21} indicate that even very sparse graph augmentations can somehow benefit the recommendation model, which suggests that CL-based recommendation may work in a way that differs from our current understanding. To better understand how CL enhances recommendation, we first investigate the necessity of graph augmentation in SGL. The original SGL paper \cite{wujc2021self} proposed three variants: SGL-ND (-ND for node dropout), SGL-ED (-ED for edge dropout), and SGL-RW (-RW for random walk, i.e., multi-layer edge dropout). To create a control group, we introduce a new variant of SGL, referred to as \textbf{SGL-WA} (-WA for without augmentation), where the CL loss is defined as follows:
\begin{equation}
	\label{loss:cl2}
	\mathcal{L}_{cl}=\sum_{i\in \mathcal{B}}-\log \frac{\exp ( 1 / \tau)}{\sum_{j \in \mathcal{B}} \exp (\mathbf{z}_{i}^{\top}\mathbf{z}_{j} / \tau)}.
\end{equation}
Because no augmentations are used in SGL-WA, we have $\mathbf{z}_{i}^{\prime}=\mathbf{z}_{i}^{\prime\prime}=\mathbf{z}_{i}$. The performance comparison is conducted on three benchmark datasets: \textit{Yelp2018} \cite{he2020lightgcn}, \textit{Amazon-Kindle} \cite{he2016ups} and \textit{Alibaba-iFashion} \cite{wujc2021self}. A 3-layer setting is adopted and the hyperparameters are tuned according to the original paper of SGL (more experimental settings can be found in Section \ref{sec:exsettings}). The results are presented in Table \ref{table:necessity} where the highest values are marked in bold type. 

\begin{table}[t]
	\scriptsize
	\caption{Performance comparison of different SGL variants.}
	\label{table:necessity}
	\renewcommand\arraystretch{1.0}
	\begin{center}
{
	\begin{tabular}{c|cc|cc|cc}
		\toprule
		\multirow{2}{*}{\textbf{Method}}&\multicolumn{2}{c}{\textbf{Yelp2018}}& \multicolumn{2}{c}{\textbf{Kindle}} & \multicolumn{2}{c}{\textbf{iFashion}} \cr
		\cmidrule(lr){2-3}\cmidrule(lr){4-5}\cmidrule(lr){6-7}&\textbf{Recall} & \textbf{NDCG}  & \textbf{Recall} & \textbf{NDCG} & \textbf{Recall} & \textbf{NDCG}  \\ \hline
		LightGCN & 0.0639 &0.0525& 0.2053 & 0.1315 & 0.0955 & 0.0461 \\
		SGL-ND &0.0644 & 0.0528 & 0.2069 & 0.1328 & 0.1032 & 0.0498 \\
		SGL-ED &\textbf{0.0675} & \textbf{0.0555} & 0.2090 & \textbf{0.1352} & 0.1093 & \textbf{0.0531} \\
		SGL-RW &0.0667 & 0.0547 & \textbf{0.2105} & 0.1351 & \textbf{0.1095} & \textbf{0.0531} \\
		SGL-WA &0.0671 & 0.0550 & 0.2084 & 0.1347 & 0.1065 & 0.0519  \\
		\bottomrule
		\end{tabular}}
	\end{center}
\end{table}

The results indicate that all graph augmentation-based variants of SGL outperform LightGCN, providing evidence of the effectiveness of CL. However, SGL-WA is also surprisingly competitive, performing on par with SGL-ED and SGL-RW, and even outperforming SGL-ND across all datasets. These results lead to two conclusions: (1) while graph augmentations do work, their effectiveness is not as significant as anticipated, and the contrastive loss InfoNCE contributes the most to the performance gains. This finding explains why even highly sparse graph augmentations can provide useful information in recent studies (e.g., \cite{yu2021socially,zhou2021selfcf,DBLP:conf/sigir/LeeKJPY21}); (2) not all graph augmentations have a positive impact, and identifying the useful ones requires an extensive trial-and-error process. Certain graph augmentations, such as node dropout, may distort the original graph by removing critical nodes (e.g., hubs) and their associated edges, resulting in disconnected subgraphs that share little learnable invariance with the original graph. On the other hand, edge dropout poses a lower risk of significantly perturbing the original graph, giving SGL-ED/RW a slight edge over SGL-WA. However, considering the cost of regularly reconstructing the adjacency matrices during training, it is reasonable to search for better alternatives.

\subsection{Uniformity Is What Really Matters}
The previous section reveals that the InfoNCE contrastive loss is crucial to CL-based recommendation. However, it is still unclear how it operates. Previous research on visual representation learning \cite{wang2020understanding} has shown that pre-training with InfoNCE intensifies two properties: feature alignment of positive pairs and feature distribution uniformity on the unit hypersphere. It is unknown whether CL-based recommendation methods exhibit similar patterns under a joint learning setting. In this study, we focus on investigating the uniformity since the goal of $\mathcal{L}_{rec}$ in recommendation is to align the interacted user-item pairs.
\par
In our preliminary study \cite{yu2022graph}, we displayed the distribution of 2,000 randomly sampled users after optimizing the InfoNCE loss. To further investigate this, in this version, we sample both users and items. We rank users and items according to their popularity and randomly sample 500 hot items from the top 5\% interactions group and 500 cold items from the bottom 80\% interactions group. We also randomly sample users in the same way. We then map the learned representations to a 2-dimensional space with t-SNE \cite{van2008visualizing} and plot the 2D feature distributions in Fig. \ref{figure:dist}. We also visualize the Gaussian kernel density estimation \cite{botev2010kernel} of $\arctan$(feature\_y/feature\_x) on the unit hypersphere $\mathcal{S}^{1}$ for a clearer presentation.
\par
From Fig. \ref{figure:dist}, we can observe a stark contrast between the features/density estimations learned by LightGCN and CL-based recommendation models. LightGCN learns highly clustered features, and the density curves have steep rises and falls. Moreover, we notice that hot users and hot items have similar distributions, and cold users cling to hot items, with only a small number of users scattered among the cold items. This biased pattern leads the model to continually expose hot items to most users, generating run-of-the-mill recommendations. We hypothesize that two issues cause this biased distribution: a fraction of items often account for most interactions in recommender systems \cite{YinCLYC12}, and the notorious over-smoothing problem \cite{chen2020measuring} that makes embeddings locally similar, thus aggravating the Matthew Effect. In contrast, the features learned by SGL variants in the second and third columns are more evenly distributed, with less sharp density estimation curves, regardless of graph augmentations. For reference, we plot the features learned only by optimizing the InfoNCE loss in SGL-ED in the fourth column. Without the effect of $\mathcal{L}_{rec}$, the features are almost subject to uniform distributions. The following inference provides a theoretical justification for this pattern. By rewriting Eq. (\ref{loss:cl}), we can derive,

\begin{equation}
	\label{eq:justification}
		\mathcal{L}_{cl}=\sum_{i \in \mathcal{B}}\Big(-\mathbf{z}_{i}^{\prime\top}\mathbf{z}_{i}^{\prime \prime} / \tau + \log\sum_{j \in \mathcal{B}} \exp (\mathbf{z}_{i}^{\prime\top}\mathbf{z}_{j}^{\prime \prime} / \tau)\Big).
\end{equation}
When the representations of different augmentations of the same node are perfectly aligned (SGL-WA is analogous to this case), we have
\begin{equation}
\mathcal{L}_{cl}=\sum_{i \in \mathcal{B}}\Bigg(-1 / \tau + \log\Big(\sum_{j \in \mathcal{B}/\{i\}} \exp (\mathbf{z}_{i}^{\prime\top}\mathbf{z}_{j}^{\prime \prime} / \tau) + \exp(1/\tau)\Big)\Bigg).
\end{equation}  
Since $1/\tau$ is a constant, optimizing the CL loss is actually towards minimizing the cosine similarity between different node representations, which will push different nodes away from each other. 
\par
Upon examining Table \ref{table:necessity} and Fig. \ref{figure:dist}, we hypothesize that the improved uniformity of the learned feature distribution is the main driver of performance gains. This uniformity can mitigate popularity bias and promote long-tail items, as discussed in Section \ref{sgl_vs_xsimgcl}, since more evenly distributed representations can better preserve the intrinsic characteristics of nodes and improve generalization. It also offers a plausible explanation for the surprisingly strong performance of SGL-WA. However, it is worth noting that the relationship between uniformity and performance is not linear. Pursuing excessive uniformity may compromise the ability of the recommendation loss to align interacted pairs and similar users/items, thereby leading to a decline in recommendation performance.

\section{Proposed Method}
\label{sec:XSimGCL}
\subsection{Noise-Based Augmentation}
Based on the findings above, we speculate that by adjusting the uniformity of the learned representation in a certain scope, contrastive recommendation models can be improved. Since manipulating the graph structure for controllable uniformity is intractable and time-consuming, we shift attention to the embedding space. Inspired by the adversarial examples \cite{goodfellow2014explaining} which are constructed through adding imperceptible perturbation to the images, we propose to directly add random noises to the representation for an efficient augmentation.
\par

\begin{figure}[t]
	\centering
	\includegraphics[width=.3\textwidth]{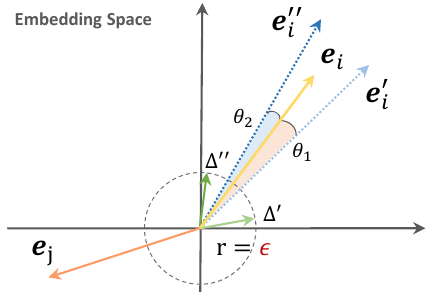}
	\caption{An illustration of the proposed random noise-based data augmentation.}
	\label{figure:aug}
	\vspace{-10pt}		
\end{figure}

Formally, given a node $i$ and its representation $\mathbf{e}_{i}$ in the $d$-dimensional embedding space, we can implement the following representation-level augmentation:
\begin{equation}
	\mathbf{e}_{i}^{\prime} = \mathbf{e}_{i} + \Delta^{\prime}_{i},\,\,\,\mathbf{e}_{i}^{\prime\prime} = \mathbf{e}_{i} + \Delta^{\prime\prime}_{i},
\end{equation}
where the added noise vectors $\Delta^{\prime}_{i}$ and $\Delta^{\prime\prime}_{i}$ are subject to $\|\Delta\|_{2}=\epsilon$ and $\epsilon$ is a small constant. This constraint of magnitude makes $\Delta$ numerically equivalent to points on a hypersphere with the radius $\epsilon$. Besides, it is required that:
\begin{equation}
	\Delta=\mathbf{\omega} \odot\sign(\mathbf{e}_{i}),	\,\,\mathbf{\omega} \in \mathbb{R}^{d}\sim U(0,1),
\end{equation}
which forces $\mathbf{e}_{i}$, $\Delta^{\prime}$ and $\Delta^{\prime\prime}$ to be in the same hyperoctant, so that adding the noises to $\mathbf{e}_{i}$ will not result in a large deviation and construct less informative augmentations of $\mathbf{e}_{i}$. Geometrically, adding these scaled noise vectors to $\mathbf{e}_{i}$ corresponds to rotating  it by two small angles ($\theta_{1}$ and $\theta_{2}$ shown in Fig. \ref{figure:aug}). This generates two augmented representations, $\mathbf{e}_{i}^{\prime}$ and $\mathbf{e}_{i}^{\prime\prime}$, that retain most of the information from the original representation while also introducing some differences. We also hope the learned representations can spread out in the entire embedding space so as to fully utilize the expression power of the space. Zhang \textit{et al.} \cite{zhang2017learning} proved that uniform distribution has such a property. We then choose to generate the noises from a uniform distribution. Though it is technically difficult to make the learned distribution approximate a uniform distribution in this way, it can statistically bring a hint of uniformity to the augmentations. 

\subsection{Simple Contrastive Recommendation Model}
\subsubsection{A Review of SimGCL}
Before presenting XSimGCL, we first briefly review SimGCL proposed in our conference paper \cite{yu2022graph} for a better understanding of the new contributions. As shown in Fig. \ref{fig:architecture}, SimGCL follows the paradigm of graph CL-based recommendation portrayed in Fig. \ref{fig:gcl}. It consists of three encoders: one is for the recommendation task and the other two are for the contrastive task. SimGCL employs LightGCN as the backbone to learn graph representations. Since LightGCN is network-parameter-free, the input user/item embeddings are the only parameters to be learned. The ordinary graph encoding is used for recommendation which follows Eq. (\ref{eq:lightgcn}) to propagate node information. Meanwhile, in the other two encoders SimGCL employs the proposed noise-based augmentation approach, and adds different uniform random noises to the aggregated embeddings at each layer to obtain perturbed representations. This noise-involved representation learning can be formulated as:
\begin{equation}
	\label{propagation}
		\mathbf{E}^{\prime}=\frac{1}{L}\sum_{l=1}^{L}(\bar{\mathbf{A}}^{l}\mathbf{E}^{(0)}+\bar{\mathbf{A}}^{l-1}\mathbf{\Delta}^{(1)}+...+\bar{\mathbf{A}}\mathbf{\Delta}^{(l-1)}+\mathbf{\Delta}^{(l)})
\end{equation} 
Note that we skip the input embedding $\mathbf{E}^{(0)}$ in all the three encoders when calculating the final representations, because we find that skipping it can lead to better performance. The possible reason is discussed in Section \ref{theoretical}. Finally, we substitute the learned representations into the joint loss presented in Eq. (\ref{joint}) then use Adam to optimize it.  

\subsubsection{XSimGCL - Simpler Than Simple}
Compared to SGL, SimGCL is much simpler because the constant graph augmentation is no longer required. However, the cumbersome architecture of SimGCL makes it less than perfect. For each computation, it requires three forward/backward passes to update the input node embeddings. Though it seems a convention to separate the pipelines of the recommendation task and the contrastive task in CL-based recommender systems \cite{wujc2021self,yu2022graph,yao2021self,xie2022contrastive}, we question the necessity of this architecture.
\par
As suggested by \cite{tian2020makes}, there is a sweet spot when using CL where the mutual information between correlated views is neither too high nor too low. In SimGCL's architecture, however, the mutual information between a pair of views of the same node could always be very high, since both embeddings contain information from $L$ hops of neighbors. Contrasting them with each other may be less effective. This is also a common problem in many CL-based recommendation models under the paradigm in Fig. \ref{fig:gcl}. To address this, we propose contrasting different layer embeddings. These embeddings share some common information but differ in aggregated neighbors and added noises, which conform to the sweet spot theory. Furthermore, since the magnitude of added noises is minuscule, we can directly use the perturbed representations for recommendation. The noises are similar to the dropout trick and are only applied during training. In the test phase, the model switches to the ordinary mode without noises. 
\par
Benefitting from this design, we can streamline the architecture of SimGCL by merging its encoding processes. This derives a new architecture that has only one-time forward/backward pass in a mini-batch computation. We name this new method \textbf{XSimGCL} and illustrate it in Fig. \ref{fig:architecture}. The perturbed representation learning of XSimGCL is as the same as that of SimGCL. The joint loss of XSimGCL is formulated as:
\begin{equation}
\begin{split}
	\mathcal{L} =& -\sum_{(u,i) \in \mathcal{B}}\log\left(\sigma(\mathbf{e}_{u}^{\prime\top}\mathbf{e}_{i}^{\prime}-\mathbf{e}_{u}^{\prime\top}\mathbf{e}^{\prime}_{j})\right)+\\ &\lambda\sum_{i \in \mathcal{B}}-\log \frac{\exp (\mathbf{z}_{i}^{\prime\top}\mathbf{z}_{i}^{l^{*}} / \tau)}{\sum_{j \in \mathcal{B}} \exp (\mathbf{z}_{i}^{\prime\top}\mathbf{z}_{j}^{l^{*}} / \tau)},	
\end{split}
\end{equation}
where $l^{*}$ denotes the layer to be contrasted with the final layer. Contrasting two intermediate layers is optional, but the experiments in Section \ref{hyper-in} show that involving the final layer leads to the optimal performance.

\begin{figure}[t]
	\centering
	\includegraphics[width=.49\textwidth]{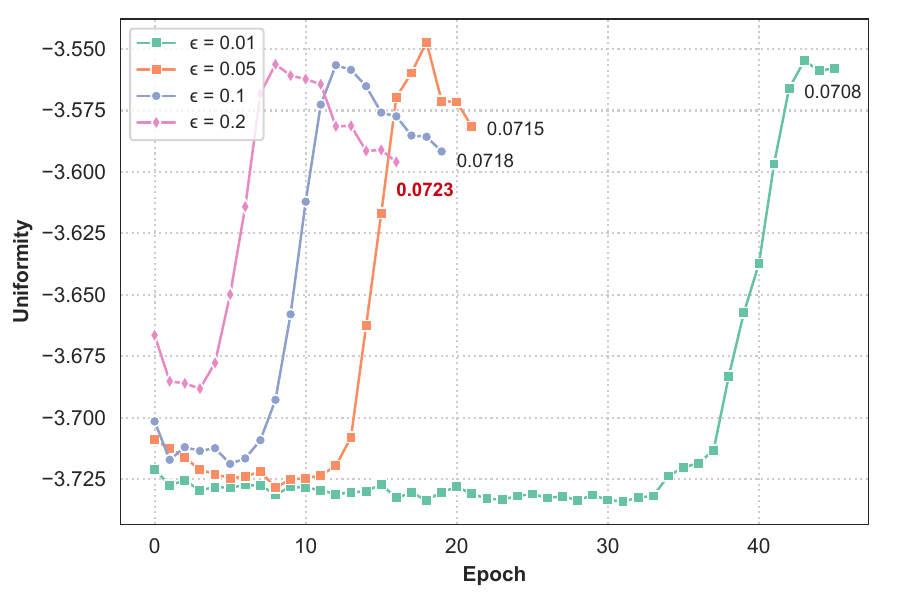}
	\caption{Trends of uniformity with different $\epsilon$. Lower values on the y-axis are better. We present the Recall@20 values of XSimGCL with different $\epsilon$ when it reaches convergence.}
	\label{fig:uniform}				
\end{figure}

\subsubsection{Ability to Adjust Uniformity Through Changing $\epsilon$}
XSimGCL enables explicit control over the deviation of augmented representations from their originals through the manipulation of $\epsilon$. Larger values of $\epsilon$ promote greater uniformity in the representation distribution, as the added noise is propagated as part of the gradients during optimization of the contrastive loss. The uniformity is then regularized toward higher levels due to the sampling of noise from a uniform distribution. To verify this claim, we conduct an experiment utilizing the logarithm of the average pairwise Gaussian potential, commonly referred to as the Radial Basis Function (RBF) kernel for measuring uniformity \cite{wang2020understanding}, which is defined as follows: 
\begin{equation}
\mathcal{L}_{\text {uniform }}(f)=\log \underset{\underset{u, v\ \sim\ p_{\text {node}}}{\scriptscriptstyle i.i.d}}{\mathbb{E}}e^{-2\|f(u)-f(v)\|_{2}^{2}}.
\label{metric}
\end{equation}
where $f(u)$ outputs the $L_{2}$ normalized embedding of $u$. 
\par
The experiment involves selecting popular items (i.e., those with more than 200 interactions) and randomly sampling 5,000 users from the Yelp2018 dataset to form user-item pairs. We then evaluate the uniformity of the representations learned by XSimGCL using Equation \ref{metric}. Using a 3-layer setting with a fixed value of $\lambda=0.2$, we tune $\epsilon$ during training to observe the effect on uniformity. We monitor the uniformity after each epoch until convergence is reached. As demonstrated in Figure \ref{fig:uniform}, our results show consistent trends across all curves. Initially, all curves display highly uniform representation distributions, likely due to the use of Xavier initialization, a special uniform distribution, to initialize the input embeddings. As training progresses, uniformity declines as a result of $\mathcal{L}_{rec}$. However, uniformity increases after reaching a minimum and continues to increase until convergence is achieved. Moreover, we find that larger values of $\epsilon$ facilitate greater uniformity in the representation distribution, which correspond to better performance. These results provide evidence for the assertion that increased uniformity can enhance performance. We also observe a correlation between the convergence speed and the magnitude of the noise, which we elaborate on in Section \ref{hyper-in}.

\subsection{Theoretical Analysis with Graph Spectrum}
\label{theoretical}
So far, our empirical findings have demonstrated the capability of XSimGCL to attain a more uniform distribution of representations. Next we theoretically reveal the necessity and effectiveness of the proposed cross-layer contrast through the lens of graph spectrum, showing it can enhance the efficacy of graph CL, by harnessing the high-frequency information in representation-level augmentations.

The graph Laplacian, denoted by $\boldsymbol{L} = \mathbf{D} - \mathbf{A}$, is a symmetric positive semi-definite matrix. 
The eigendecomposition of $\boldsymbol{L}$ yields orthonormal eigenvectors in $\mathbf{U} \in \mathbb{R}^{n \times n}$ and a diagonal matrix of eigenvalues, $\boldsymbol{\Lambda} = \text{diag}(\lambda_1, \dots, \lambda_n)$. The graph's Fourier transform can be defined using this eigendecomposition, where each eigenvector corresponds to a Fourier mode, and each eigenvalue corresponds to a frequency of the graph, implying amplitudes of different frequency components. The set of eigenvalues is referred to as graph spectrum. Suppose that $\mathbf{x} \in \mathbb{R}^n$ is a signal defined on the graph's vertices (i.e., node embedding in our method). In that case, the graph Fourier transform of $\mathbf{x}$ is defined as $\hat{\mathbf{x}} = \mathbf{U}^{\top}\mathbf{x}$, with its inverse operation given by $\mathbf{x} = \mathbf{U} \hat{\mathbf{x}}$ \cite{kipf2016semi}. This definition enables us to perform graph convolution between signal $\mathbf{x}$ and filter $\mathbf{g}$ in the spectral domain as:
\begin{equation} \label{eq:spectral_conv_definition}
\mathbf{g} * \mathbf{x} = \mathbf{U} \left( (\mathbf{U}^\top \mathbf{g}) \odot (\mathbf{U}^\top \mathbf{x}) \right) = \mathbf{U} \hat{\mathbf{G}} \mathbf{U}^\top \mathbf{x},
\end{equation}
where $\odot$ denotes element-wise multiplication, and $\hat{\mathbf{G}} = \text{diag}\left(g_{\theta}(\lambda_1), \dots, g_{\theta}(\lambda_n)\right)$ denotes a diagonal matrix with spectral filter coefficients on the diagonal, where $g_{\theta}$ is a function of the eigenvalues of $\boldsymbol{L}$.

As it is often time-consuming to decompose the Laplacian matrix $\boldsymbol{L}$ to acquire $\mathbf{U}$, Kipf \textit{et al.} \cite{kipf2016semi} proposed to approximate this graph convolution with the first-order Chebyshev polynomials, deriving:
\begin{equation}
	\label{cheby}
	\mathbf{g} * \mathbf{x}=\left(\mathbf{I}+\mathbf{D}^{-1 / 2} \mathbf{A D}^{-1 / 2}\right) \mathbf{x}.
\end{equation}
Since the normalized Laplacian matrix $\boldsymbol{L}_{\mathrm{sym}}$ = $\mathbf{I}-\mathbf{D}^{-1 / 2} \mathbf{A D}^{-1 / 2}$, then $\mathbf{I}+\mathbf{D}^{-1 / 2} \mathbf{A D}^{-1 / 2}$ = $2\mathbf{I}-\boldsymbol{L}_{\mathrm{sym}}$. Hence, Eq (\ref{cheby}) can be re-written to:
\begin{equation}
\mathbf{g} * \mathbf{x}=\mathbf{U} (2\mathbf{I}-\boldsymbol{\Lambda}) \mathbf{U}^\top \mathbf{x}.
\end{equation}
Here $\mathbf{U}$ and $\mathbf{\Lambda}$ refer to the eigenvectors and eigenvalues ($\lambda_i \in [0,2]$) of $\boldsymbol{L}_{\mathrm{sym}}$. When stacking $K$ graph convolutional layers, we can obtain the convolutional kernel $(2\mathbf{I}-\boldsymbol{\Lambda})^{K}$. Apparently, this kernel leads to heavily shrunk filter coefficients at frequencies $\lambda_i>1$ and over-amplified filter coefficients at frequencies $\lambda_i<1$. It results in a low-pass filter where the high-frequency information is attenuated. As for LightGCN, the backbone of XSimGCL, it discards the self-loop in the adjacency matrix, leading to a graph convolution defined as:
\begin{equation}
	\mathbf{g} * \mathbf{x}=\mathbf{D}^{-1 / 2} \mathbf{A D}^{-1 / 2}\mathbf{x}=\mathbf{U} (\mathbf{I}-\boldsymbol{\Lambda}) \mathbf{U}^\top \mathbf{x}.
	\end{equation}
With the convolutional kernel of LightGCN, both the low-frequency information and high-frequency information can pass while the odd powers of ($\mathbf{I}-\boldsymbol{\Lambda}$) yield negative filter coefficients at frequencies $\lambda_i>1$. 

As demonstrated in a recent work \cite{liu2022revisiting}, a general rule for selecting effective augmentations is: ``the difference of the high-frequency parts between two augmentations should be larger than that of low-frequency parts". Since the convolutional kernel of LightGCN can alternately generate positive and negative filter coefficients for high-frequency components with the increase of $K$, using the cross-layer contrast instead of the final-layer contrast coincides with this rule because the difference of the high-frequency parts from different layers is larger than that from the final layers as well as the difference of the low-frequency parts. One major piece of evidence for the utilization of high-frequency information in cross-layer contrast is that, when we add the self-loop, namely, using Eq (14) to propagate features, a drastic performance drop is observed as the high-frequency information is attenuated. Based on this, we can derive an explanation for the performance drop led by integrating $\mathbf{E}^{(0)}$ into Eq. (\ref{propagation}). When $\mathbf{E}^{(0)}$ is included, for a one-layer LightGCN, $\mathbf{E}'=\mathbf{E}^{(0)}+\bar{\mathbf{A}}\mathbf{E}^{(0)}+\mathbf{\Delta}^{(0)}=(\mathbf{I}+\bar{\mathbf{A}})\mathbf{E}^{(0)}+\mathbf{\Delta}^{(0)}=(\mathbf{I}+\mathbf{D}^{-1 / 2} \mathbf{A D}^{-1 / 2})\mathbf{E}^{(0)}+\mathbf{\Delta}^{(0)}$, which is equivalent to propagating with Eq. (\ref{cheby}) at the first layer.
\subsection{Complexity}
In this section, we analyze the theoretical complexity of XSimGCL and compare it with LightGCN, SGL-ED and its predecessor SimGCL. The discussion is within the scope of a single batch since the in-batch negative sampling is a widely used trick in CL \cite{chen2020simple}. Here we let $|A|$ be the edge number in the user-item bipartite graph, $d$ be the embedding dimension, $B$ denote the batch size, $M$ represent the node number in a batch, $L$ be the layer number, and $\rho$ denote the edge keep rate in SGL-ED. We can derive:

\begin{table}[h]
	\scriptsize
	\caption{The comparison of time complexity}
	\begin{adjustbox}{width=\columnwidth,center}
	\begin{tabular}{c|c|c|c|c}
	\hline	
	& LightGCN          & SGL-ED            & SimGCL     & XSimGCL         \\ \hline\hline
	\makecell{Adjacency\\Matrix} & \makecell{$\mathcal{O}(2|A|)$} & \makecell{$\mathcal{O}(2|A|$+$4\rho|A|)$} & \makecell{$\mathcal{O}(2|A|)$} & \makecell{$\mathcal{O}(2|A|)$}\\ \hline
    \makecell{Graph\\Encoding} & \makecell{$\mathcal{O}(2|A|Ld)$} & \makecell{$\mathcal{O}((2$+$4\rho)|A|Ld)$} & \makecell{$\mathcal{O}(6|A|Ld)$} & \makecell{$\mathcal{O}(2|A|Ld)$} \\ \hline	
	Prediction & $\mathcal{O}(2Bd)$ & $\mathcal{O}(2Bd)$ & $\mathcal{O}(2Bd)$ & $\mathcal{O}(2Bd)$ \\ \hline

    Contrast & - & $\mathcal{O}(BMd)$ & $\mathcal{O}(BMd)$ & $\mathcal{O}(BMd)$\\  \hline
	\end{tabular}
\end{adjustbox}
\end{table}

\begin{itemize}[leftmargin=*]
	\item Since LightGCN, SimGCL and XSimGCL do not need graph augmentations, they only construct the normalized adjacency matrix which has $2|A|$ non-zero elements. For SGL-ED, two graph augmentations are used and each has 2$\rho|A|$ non-zero elements in their adjacency matrices.
	\item In the phase of graph encoding, a three-encoder architecture is adopted in both SGL-ED and SimGCL to learn two different augmentations, so the encoding expense of SGL-ED and SimGCL are almost three times that of LightGCN. In contrast, the encoding expense of XSimGCL is as the same as that of LightGCN.
	\item As for the prediction, all methods are trained with the BPR loss and each batch contains $B$ interactions, so they have exactly the same time cost in this regard.
	\item The computational cost of CL comes from the contrast between the positive/negative samples, which are $\mathcal{O}(Bd)$ and $\mathcal{O}(BMd)$, respectively, because each node regards the views of itself as the positives and views of other nodes as the negatives. For the sake of brevity, we mark it as $\mathcal{O}(BMd)$ since $M\gg 1$.
\end{itemize}

In these four models, SGL-ED and SimGCL are obviously the two with the highest computation costs. SimGCL needs more time in the graph encoding but SGL-ED requires constant graph augmentations. Since this part is usually performed on CPUs, which brings SGL-ED more expense of time in practice. By comparison, XSimGCL needs neither graph augmentations nor extra encoders. Without considering the computation for the contrastive task, XSimGCL is theoretically as lightweight as LightGCN and only spends one-third of SimGCL's training expense in graph encoding. When the actual number of epochs for training is considered, XSimGCL will show more efficiency beyond what we can observe from this theoretical analysis. 

\begin{table}[!htb]
	\renewcommand\arraystretch{1.0}
	\caption{Dataset Statistics}
	\label{Table:dataset}
	\centering
	\resizebox{\columnwidth}{!}{
		\begin{tabular}{c|cccc}
			\hline
			Dataset&\#User & \#Item &  \#Feedback  & Density\\ \hline
			\hline
			Yelp2018 &31,668 &  38,048 & 1,561,406   & 0.13\%\\
			Amazon-Kindle & 138,333 & 98,572 & 1,909,965 & 0.014\%\\
			Alibaba-iFashion&300,000  &81,614 & 1,607,813 & 0.007\%\\
			Amazon-Electronics & 719,376 & 159,364 & 5,460,975 & 0.005\%\\
			\hline
		\end{tabular}
}
\end{table}

\begin{table*}[t]
	\footnotesize
	\caption{Performance Comparison for different CL methods on three benchmarks.}
	\label{Table:comparison}
	\renewcommand\arraystretch{1.1}
	\begin{center}
{	
	\begin{tabular}{cccccccccc}
		\toprule
		\multicolumn{2}{c}{\multirow{2}{*}{\textbf{Method}}}&
		\multicolumn{2}{c}{\textbf{Yelp2018}} & \multicolumn{2}{c}{\textbf{Amazon-Kindle}} & \multicolumn{2}{c}{\textbf{Alibaba-iFashion}} & \multicolumn{2}{c}{\textbf{Amazon-Electronics}}\cr
		\cmidrule(lr){3-4}\cmidrule(lr){5-6}\cmidrule(lr){7-8} \cmidrule(lr){9-10}&& \textbf{Recall@20} & \textbf{NDCG@20} &  \textbf{Recall@20} & \textbf{NDCG@20} & \textbf{Recall@20} & \textbf{NDCG@20} & \textbf{Recall@20} & \textbf{NDCG@20} \\ \hline				
		
		\multirow{6}{*}{\textbf{1-Layer}} 
		&LightGCN &0.0590 & 0.0484 & 0.1871 & 0.1186 & 0.0845 & 0.0390 & 0.0497 & 0.0298  \\
	
		&SGL-ND &0.0643  & 0.0529  & 0.1880 & 0.1192 & 0.0896 & 0.0432 & 0.0621 & 0.0459 \\
		&SGL-ED &0.0637 & 0.0526 & 0.1936 & 0.1231  & 0.0932 & 0.0447  & 0.0636 & 0.0464 \\
		&SGL-RW &0.0637 & 0.0526 & 0.1936 & 0.1231  & 0.0932 & 0.0447  & 0.0636 & 0.0464 \\
		&SGL-WA &0.0628 & 0.0525 & 0.1918 & 0.1221  & 0.0913 & 0.0440 & 0.0631 & 0.0465 \\

		&{\textbf{SimGCL}}  &\underline{0.0689} & \underline{0.0572} & \textbf{0.2087} & \textbf{0.1361} & \underline{0.1036}&  \underline{0.0505} & \underline{0.0665} & \underline{0.0486} \\
		&{\textbf{XSimGCL}} &\textbf{0.0692} & \textbf{0.0582} & \underline{0.2071} & \underline{0.1339} & \textbf{0.1069} & \textbf{0.0527}  & \textbf{0.0690} & \textbf{0.0500}  \\
	    \midrule[0.75pt]

		\multirow{7}{*}{\textbf{2-Layer}} 
		&LightGCN &0.0622 & 0.0504 & 0.2033 & 0.1284 & 0.1053 & 0.0505 & 0.0545 & 0.0352  \\

		&SGL-ND &0.0658 & 0.0538 & 0.2020 & 0.1307 & 0.0993 & 0.0484 & 0.0665 & 0.0465 \\
		&SGL-ED &0.0668 & 0.0549 & 0.2084 & 0.1341 & 0.1062 & 0.0514  & 0.0688 & 0.0496 \\
		&SGL-RW &0.0644 & 0.0530 & \underline{0.2088} & \underline{0.1345} & 0.1053 & 0.0512  & 0.0692 & \underline{0.0497} \\
		&SGL-WA &0.0653 & 0.0544 & 0.2068 & 0.1330 & 0.1028 & 0.0501  & 0.0681 & 0.0489 \\
	
		&{\textbf{SimGCL}} & \underline{0.0719}  & \underline{0.0601} & 0.2071 & 0.1341 & \underline{0.1119} & \underline{0.0548}   & \underline{0.0698} & 0.0493 \\
		&{\textbf{XSimGCL}} &\textbf{0.0722} & \textbf{0.0604} & \textbf{0.2114}  & \textbf{0.1382} & \textbf{0.1143} & \textbf{0.0559} & \textbf{0.0704}  & \textbf{0.0521} \\

	    \midrule[0.75pt]	
		\multirow{7}{*}{\textbf{3-Layer}} 
		&LightGCN &0.0639 & 0.0525 & 0.2057 & 0.1315 & 0.0955 & 0.0461 & 0.0544 & 0.0341 \\

		&SGL-ND &0.0644 & 0.0528 & 0.2069 & 0.1328 & 0.1032 & 0.0498 & 0.0681 & 0.0475 \\
		&SGL-ED &0.0675 & 0.0555& 0.2090  & 0.1352 & 0.1093 & 0.0531 & 0.0704 & 0.0486 \\
		&SGL-RW &0.0667 & 0.0547& \underline{0.2105} & 0.1351 & 0.1095 & 0.0531 & 0.0702 & 0.0487 \\
		&SGL-WA &0.0671 & 0.0550& 0.2084 & 0.1347 & 0.1065 & 0.0519 & 0.0694 & \underline{0.0496} \\

		&{\textbf{SimGCL}} &\underline{0.0721} & \underline{0.0601} & 0.2104 & \underline{0.1374} & \underline{0.1151} & \underline{0.0567}  & \underline{0.0715}  &  0.0492 \\
		&{\textbf{XSimGCL}} &\textbf{0.0723} & \textbf{0.0604} & \textbf{0.2147} & \textbf{0.1415} & \textbf{0.1196} & \textbf{0.0586} &  \textbf{0.0750} & \textbf{0.0531}  \\	
		
		\midrule[0.75pt]	
		\multirow{7}{*}{\textbf{4-Layer}} 
		&LightGCN &0.0619 & 0.0505 & 0.1954 & 0.1247 & 0.0918 & 0.0427 & 0.0560 &  0.0354 \\

		&SGL-ND &0.0639 & 0.0526 & 0.2042 & 0.1301 & 0.1040 & 0.0496 & 0.0681 & 0.0475	 \\
		&SGL-ED &0.0673 & 0.0553 & 0.2082 & 0.1315 & 0.1127 & 0.0540  & 0.0705 & 0.0488 \\
		&SGL-RW &0.0674 & 0.0553 & 0.2074 & 0.1314 & 0.1126 & 0.0540  & 0.0711 & 0.0490 \\
		&SGL-WA &0.0671 & 0.0550 & 0.2067 & 0.1312 & 0.1111& 0.0533  & 0.0707 & 0.0487 \\

		&{\textbf{SimGCL}} &\underline{0.0726} & \underline{0.0604} & \underline{0.2102} & \underline{0.1365} & \underline{0.1170} & \underline{0.0572}  & \underline{0.0723} &  \underline{0.0512} \\
		&{\textbf{XSimGCL}} &\textbf{0.0733} & \textbf{0.0606} & \textbf{0.2135} & \textbf{0.1401} & \textbf{0.1205} & \textbf{0.0582} & \textbf{0.0747} &  \textbf{0.0533} \\	

		\midrule[0.75pt]	
		\multirow{7}{*}{\textbf{5-Layer}} 
		&LightGCN &0.0610 & 0.0501 & 0.1965 & 0.1260 & 0.0930 & 0.0433 & 0.0560 & 0.0347  \\

		&SGL-ND &0.0636 & 0.0524 & 0.2049 & 0.1307 & 0.1025 & 0.0487 & 0.0685 & 0.0473 \\
		&SGL-ED &0.0678 & 0.0557 & 0.2088 & 0.1322 & 0.1133 & 0.0543  & 0.0709 & 0.0492 \\
		&SGL-RW &0.0677 & 0.0555 & 0.2112 & 0.1344 & 0.1126 & 0.0540  & 0.0707 & 0.0492 \\
		&SGL-WA &0.0676 & 0.0554 & 0.2076 & 0.1316 & 0.1116 & 0.0537  & 0.0702 & 0.0488 \\

		&{\textbf{SimGCL}} &\underline{0.0722} & \underline{0.0599} & \underline{0.2118} & \underline{0.1376} & \underline{0.1180} & \underline{0.0571} & \underline{0.0726} & \underline{0.0510}  \\
		&{\textbf{XSimGCL}} &\textbf{0.0729} & \textbf{0.0602} & \textbf{0.2134} & \textbf{0.1393} & \textbf{0.1202} & \textbf{0.0580} & \textbf{0.0749} & \textbf{0.0526}  \\	
		\bottomrule
		\end{tabular}}
	\end{center}
\end{table*}

\section{Experiments}
\label{sec:experiments}
\subsection{Experimental Settings}
\label{sec:exsettings}
\noindent\textbf{Datasets.} For reliable and convincing results, we conduct experiments on four public large-scale datasets: Yelp2018 \cite{he2020lightgcn}, Amazon-kindle \cite{wujc2021self}  Alibaba-iFashion \cite{wujc2021self} and Amazon-Electronics \cite{ni2019justifying} to evaluate XSimGCL/SimGCL. The statistics of these datasets are presented in Table \ref{Table:dataset}. We split the datasets into three parts (training set, validation set, and test set) with a 7:1:2 ratio. Following \cite{wujc2021self,he2020lightgcn}, we first search the best hyperparameters on the validation set, and then we merge the training set and the validation set to train the model and evaluate it on the test set where the relevancy-based metric Recall@$20$ and the ranking-aware metric NDCG@$20$ are used. For a rigorous and unbiased evaluation, the reported result are the average values of 5 runs, with all the items being ranked. \par
\noindent\textbf{Baselines.} Besides LightGCN and the SGL variants, the following recent data augmentation-based/CL-based recommendation models are compared.
\begin{itemize}[leftmargin=*]
	\item \textbf{DNN+SSL} \cite{yao2021self} is a recent DNN-based recommendation method which adopts the similar architecture in Fig. \ref{fig:gcl}, and conducts feature masking for CL. 
	\item \textbf{BUIR} \cite{DBLP:conf/sigir/LeeKJPY21} has a two-branch architecture which consists of a target network and an online network, and only uses positive examples for self-supervised recommendation.
	\item \textbf{MixGCL} \cite{huang2021mixgcf} designs the hop mixing technique to synthesize hard negatives for graph collaborative filtering by embedding interpolation.
	\item \textbf{NCL} \cite{lin2022improving} is a very recent contrastive model which designs a prototypical contrastive objective to capture the correlations between a user/item and its context.
\end{itemize} 
\noindent\textbf{Hyperparameters.} For a fair comparison, we referred to the best hyperparameter settings reported in the original papers of the baselines and then fine-tuned them with the grid search. As for the general settings, we create the user and item embeddings with the Xavier initialization of dimension 64; we use Adam to optimize all the models with the learning rate 0.001; the $L_{2}$ regularization coefficient 10$^{-4}$ and the batch size 2048 are used, which are common in many papers \cite{he2020lightgcn,wujc2021self,wang2020disentangled}. In SimGCL, XSimGCL and SGL, we empirically let the temperature $\tau=0.2$ because this value is often reported a great choice in papers on CL \cite{wujc2021self,wang2020understanding}. An exception is that we let $\tau=0.15$ for XSimGCL on Yelp2018, which brings a slightly better performance. Note that although the paper of SGL \cite{wujc2021self} uses Yelp2018 and Alibaba-iFashion as well, we cannot reproduce their results on Alibaba-iFashion with their given hyperparameters under the same experimental setting. So we re-search the hyperparameters of SGL and choose to present our results on this dataset in Table \ref{Table:comparison}.
\par

\begin{table}[!htb]
	\renewcommand\arraystretch{1.0}
	\caption{The best hyperparameters of compared methods.}
	\label{table:hyper}
	\centering
	\resizebox{\columnwidth}{!}{
		\begin{tabular}{c|c|c|c|c}
			\hline
			\textbf{Dataset} & \textbf{Yelp2018} & \textbf{Kindle} &  \textbf{iFashion} & \textbf{Electronics} \\ \hline
			\hline
			SGL &$\lambda$=0.1, $\rho$=0.1 & $\lambda$=0.05, $\rho$=0.1 & $\lambda$=0.05, $\rho$=0.2& $\lambda$=0.1, $\rho$=0.1\\
			\hline
			SimGCL &$\lambda$=0.5, $\epsilon$=0.1 & $\lambda$=0.1, $\epsilon$=0.1 & $\lambda$=0.05, $\epsilon$=0.1 &$\lambda$=0.2, $\epsilon$=0.1\\
			\hline
			XSimGCL &\makecell{$\lambda$=0.2, $\epsilon$=0.2,\\ $l^{*}$=2} & \makecell{$\lambda$=0.2, $\epsilon$=0.1,\\ $l^{*}$=1} & \makecell{$\lambda$=0.05, $\epsilon$=0.05, \\$l^{*}$=4} &\makecell{$\lambda$=0.2, $\epsilon$=0.1, \\$l^{*}$=3}\\
			\hline
		\end{tabular}
}
\end{table}

\begin{table*}[t]
	\caption{Performance comparison with other models.}
	\footnotesize
	\label{Table:sota}
	\renewcommand\arraystretch{1.0}
	\begin{center}
{
	\begin{tabular}{ccccccccc}
		\toprule
		\multirow{2}{*}{\textbf{Method}}&\multicolumn{2}{c}{\textbf{Yelp2018}}& \multicolumn{2}{c}{\textbf{Amazon-Kindle}} & \multicolumn{2}{c}{\textbf{Alibaba-iFashion}} & \multicolumn{2}{c}{\textbf{Amazon-Electronics}} \cr
		\cmidrule(lr){2-3}\cmidrule(lr){4-5}\cmidrule(lr){6-7}\cmidrule(lr){8-9}&\textbf{Recall@20} & \textbf{NDCG@20}  & \textbf{Recall@20} & \textbf{NDCG@20} & \textbf{Recall@20} & \textbf{NDCG@20} &\textbf{Recall@20} & \textbf{NDCG@20}  \\ \hline
		LightGCN  &0.0639&0.0525& 0.2057& 0.1315 & 0.1053 & 0.0505 & 0.0560 &  0.0354	\\
		NCL &  0.0682  & 0.0573  & 0.2100 & 0.1357 & 0.1132 & 0.0547 & \textit{OOM} & \textit{OOM} \\
		BUIR  & 0.0487  & 0.0404  & 0.0922 & 0.0528 & 0.0830& 0.0384 & 0.0436 & 0.0268\\	
		DNN+SSL & 0.0483  & 0.0382  & 0.1520 & 0.0989 & 0.0818 & 0.0375 & 0.0405 & 0.0238\\
		MixGCF &  0.0713  & 0.0589  & 0.2128 & 0.1327 & 0.1124 & 0.0549 & 0.0705 &0.0476\\		
		SimGCL &\underline{0.0726} & \underline{0.0604}& \underline{0.2118} & \underline{0.1376} & \underline{0.1180} & \underline{0.0571} &  \underline{0.0723} &  \underline{0.0512} \\			
		XSimGCL &\textbf{0.0733} & \textbf{0.0606}& \textbf{0.2147} & \textbf{0.1415} & \textbf{0.1205} & \textbf{0.0582} &  \textbf{0.0750} & \textbf{0.0531} \\
		\bottomrule
		\end{tabular}}
	\end{center}
\end{table*}

\subsection{SGL \textit{vs.} XSimGCL: A Comprehensive Perspective}
\label{sgl_vs_xsimgcl}
In this part, we compare XSimGCL with SGL in a comprehensive way. The experiments focus on three important aspects: recommendation performance, training time, and the ability to promote long-tail items.
\subsubsection{Performance Comparison}
We first present the performance comparison of SGL and XSimGCL/SimGCL with varying numbers of layers. We provide the best hyperparameters for each approach in Table \ref{table:hyper} to facilitate the reproducibility of our findings. Bold and underlined figures are used to denote the best and runner-up performance, respectively, with respect to the baseline method LightGCN. We note that the final layer of 1-layer XSimGCL is compared against itself. Based on the comparison results presented in Table \ref{Table:comparison}, we make the following observations:
\begin{itemize}[leftmargin=*]
\item In the majority of cases, the SGL variants, SimGCL and XSimGCL, demonstrate significant performance improvements over LightGCN. The largest performance gains are observed on the largest and sparsest dataset, Amazon-Electronics, where XSimGCL achieves a Recall@20 improvement of 33.4\% and a NDCG@20 improvement of 50.6\% compared to LightGCN, under the 4-layer setting.
\item SGL-ED and SGL-RW exhibit similar performances, both of which outperform SGL-ND by a large margin. While SGL-WA demonstrates some advantages over SGL-ND, it still falls behind SGL-ED and SGL-RW. These findings further corroborate that the InfoNCE loss is the primary factor which accounts for the performance gains, whereas heuristic graph augmentations are not as effective as expected and can even degrade the performance.
\item XSimGCL/SimGCL show the best/second best performance in almost all the cases, which demonstrates the effectiveness of the noised-based data augmentation. Particularly, on the sparser dataset - Alibaba-iFashion, they significantly outperforms the SGL variants. Additionally, it is undoubted that the evolution from SimGCL to XSimGCL is successful, bringing non-negligible performance gains.  
\item In most cases, the compared methods achieve their best performance under the 3-layer or 4-layer settings. With the models going deeper, the performance gains diminish. Notably, the performance of LightGCN decreases on three datasets, while the performance of CL-based methods remains relatively stable, which suggests that CL can mitigate the over-smoothing issue as it leads to more evenly distributed representation learning.  
\end{itemize}
\par
To further demonstrate XSimGCL's outstanding performance, we also compare it with several recent augmentation-based and CL-based recommendation models. The implementations of these methods are available in our GitHub repository SELFRec as well. According to Table \ref{Table:sota}, XSimGCL and SimGCL still outperform other methods with a great lead, achieving the best and the second best performance, respectively. NCL and MixGCF, which employ LightGCN as their backbones, also show their competitiveness. By contrast, DNN+SSL and BUIR are not as powerful as expected and even not comparable to LightGCN. We attribute their failure to: (1). DNNs are proved effective when abundant user/item features are provided. In our datasets, features are unavailable and the self-supervision signals are created by masking item embeddings, so it cannot fulfill its potential in this situation. (2). In the paper of BUIR, the authors removed long-tail users and items to guarantee a good result, but we use all the data. We also notice that BUIR performs very well on suggesting popular items but poorly on long-tail items. This may explain why the original paper uses a biased experimental setting.

\subsubsection{Comparison of Training Efficiency}
\label{sec:speed}
As has been claimed, XSimGCL is almost as lightweight as LightGCN in theory. In this part, we report the actual training time, which is more informative than the theoretical analysis. The reported figures are collected on a workstation with an Intel(R) Xeon(R) Gold 5122 CPU and a GeForce RTX 2080Ti GPU. These methods are implemented with Tensorflow 1.14, and a 2-layer setting is applied to all. \par
\begin{figure}[t]
	\centering
	\captionsetup{justification=centering}
	\includegraphics[width=.48\textwidth]{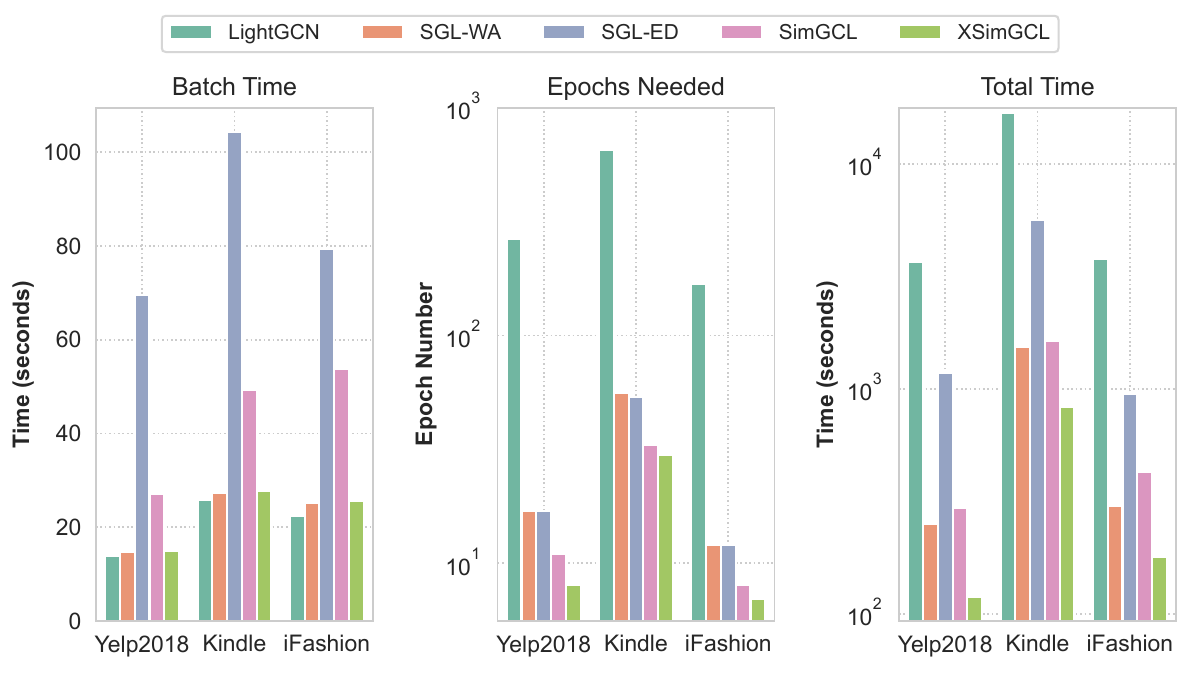}
	\caption{The training speed of compared methods. }
	\label{fig:speed}				
\end{figure}

According to Fig. \ref{fig:speed}, we have the following observations: 
\begin{itemize}[leftmargin=*]
	\item SGL-ED takes the longest time to finish the computation in a single batch, which is almost four times that of LightGCN on all the datasets. SimGCL ranks second due to its three-encoder architecture, which is almost two times that of LightGCN. Since SGL-WA, XSimGCL and LightGCN have the same architecture, there training costs for a batch are very close. The former two need a bit of extra time for the contrastive task. 
	\item LightGCN is trained with hundreds of epochs, which is at least an order of magnitude more than the epochs that other methods need. By contrast, XSimGCL needs the fewest epochs to reach convergence and its predecessor SimGCL falls behind by several epochs. SGL-WA and SGL-ED require the same number of epochs to get converged and are slower than SimGCL. When it comes to the total training time, LightGCN is still the method trained with the longest time, followed by SGL-ED and SimGCL. Due to the simple architecture, SGL-WA and XSimGCL are the last two but XSimGCL only needs about half of the cost SGL-WA spends in total.  
\end{itemize}
With these observations, we can easily draw some conclusions. First, CL can tremendously accelerate the training. Second, graph augmentations cannot contribute to the training efficiency. Third, the cross-layer contrasts not only brings performance improvement but also leads to faster convergence. By analyzing the gradients from the CL loss, we find that the noises in XSimGCL and SimGCL will add an small increment to the gradients, which works like a momentum and can explain the speedup. Compared with the final-layer contrast, the cross-layer has shorter route for gradient propagation. This can explain why XSimGCL needs fewer epochs compared with SimGCL.

\begin{figure}[t]
	\centering
	\captionsetup{justification=centering}
	\includegraphics[width=.48\textwidth]{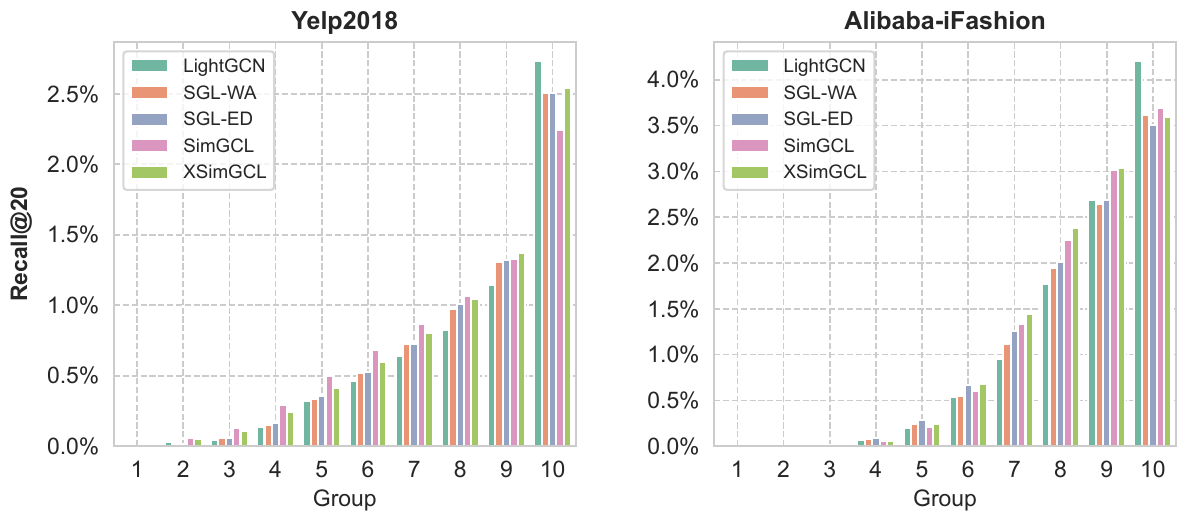}
	\caption{The ability to promote long-tail items. }
	\label{fig:debias}				
\end{figure}

\subsubsection{Comparison of Ability to Promote Long-tail Items}
Optimizing the InfoNCE loss is found to learn more evenly distributed representations, which is supposed to alleviate the popularity bias. To verify that XSimGCL upgrades this ability with the noise-based augmentation, we divided the test set into ten groups with IDs ranging from 1 to 10, each containing the same number of interactions. The higher the group ID, the more popular items it included. We then evaluated the recall@20 value of each group using a 2-layer setting, as shown in Fig. \ref{fig:debias}. 
\par
According to Fig. \ref{fig:debias}, LightGCN is inclined to recommend popular items and achieves the highest recall value on the last group. By contrast, XSimGCL and SimGCL do not show outstanding performance on group 10, but they have distinct advantages over LightGCN on other groups. Particularly, SimGCL is the standout on Yelp2018 and XSimGCL keeps strong on iFashion. Their extraordinary performance in recommending long-tail items largely compensates for their loss on the popular item group. As for the SGL variants, they fall between LightGCN and SimGCL on exploring long-tail items and exhibit similar recommendation performance on Yelp2018. SGL-ED shows a slight advantage over SGL-WA on iFashion. Combining Fig. \ref{figure:dist} with Fig. \ref{fig:debias}, we can easily find that the ability to promote long-tail items seems to positively correlate with the uniformity of representations. Since a good recommender system should suggest items that are most pertinent to a particular user instead of recommending popular items that might have been known, SimGCL and XSimGCL significantly outperforms other methods in this regard.

\subsection{Hyperparameter Investigation}
\label{hyper-in}
XSimGCL has three important hyperparameters: $\lambda$ - the coefficient of the contrastive task, $\epsilon$ - the magnitude of added noises, and $l^{*}$ - the layer to be contrasted. In this part, we investigate the model's sensitivity to these hyperparameters.
\subsubsection{Influence of $\lambda$ and $\epsilon$} 
\begin{figure}[t]
	\centering
	\captionsetup{justification=centering}
	\includegraphics[width=.48\textwidth]{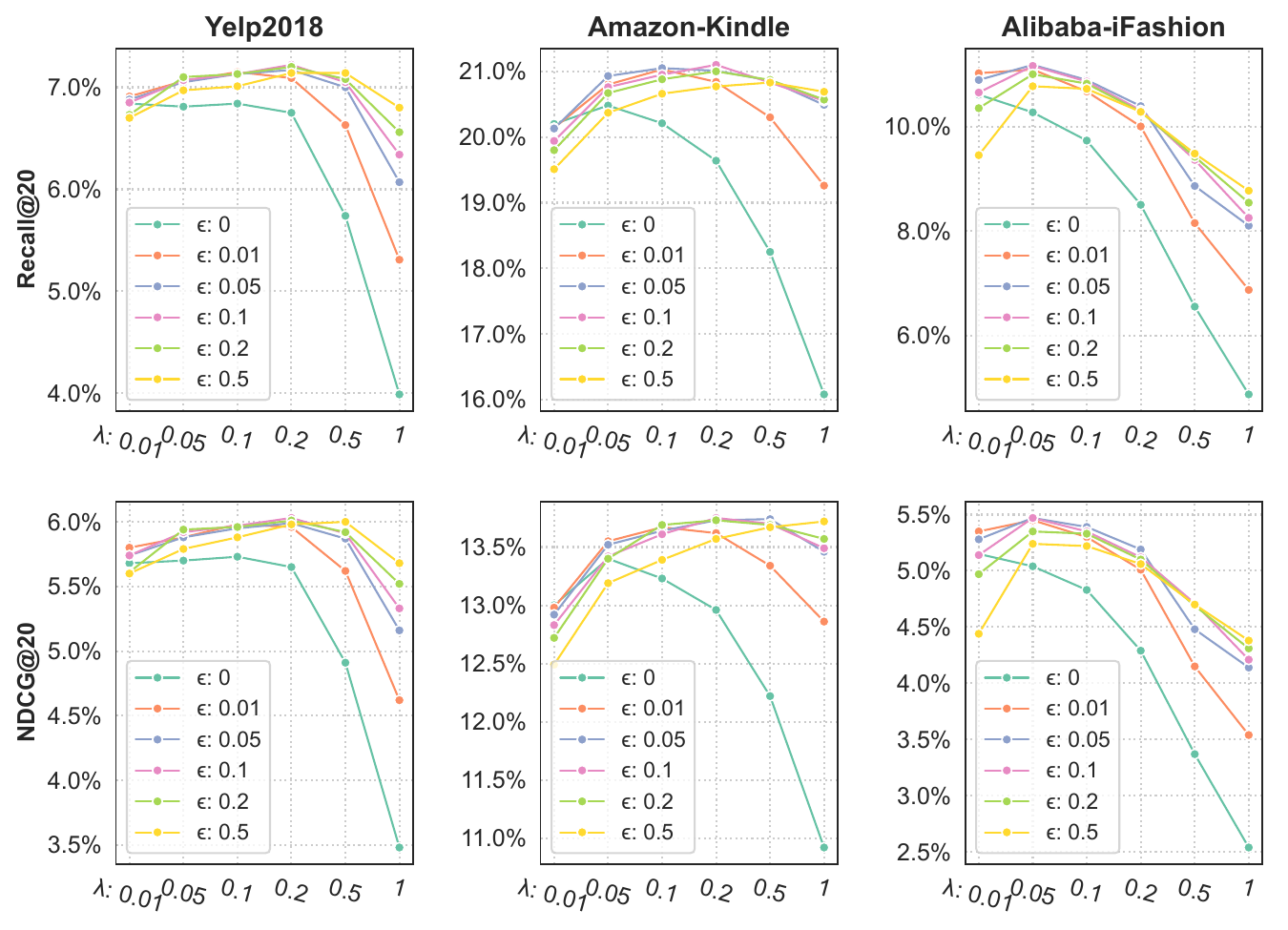}
	\caption{The influence of $\lambda$ and $\epsilon$. }
	\label{fig:sensitivity}				
\end{figure}

We perform experiments with different combinations of $\lambda$ and $\epsilon$ using the set [0.01, 0.05, 0.1, 0.2, 0.5, 1] for $\lambda$ and [0, 0.01, 0.05, 0.1, 0.2, 0.5] for $\epsilon$. We fix $l^{*}$=1 and conduct experiments with a 2-layer setting. However, we find that the best values of these two hyperparameters are also applicable to other settings. As shown in Fig. \ref{fig:sensitivity}, XSimGCL achieves its best performance on all datasets when $\epsilon$ is in the range of [0.05, 0.1, 0.2]. Without the added noise ($\epsilon$=0), we observe a significant drop in performance. When $\epsilon$ is too small (0.01) or too large (0.5), the performance also declines. A similar trend is observed when changing the value of $\lambda$. The performance is at its peak when $\lambda$=0.2 on Yelp2018, $\lambda=0.2$ on Amazon-Kindle, and $\lambda=0.05$ on Alibaba-iFashion. Our experience suggests that XSimGCL is more sensitive to changes in $\lambda$, and $\epsilon=0.1$ is usually a good and safe choice on most datasets. Moreover, we find that a larger $\epsilon$ leads to faster convergence. However, when it is too large (e.g., greater than 1), it acts like a large learning rate and causes the progressive \textit{zigzag} optimization, which overshoots the minimum.

\begin{figure}[t]
	\centering
	\captionsetup{justification=centering}
	\includegraphics[width=.48\textwidth]{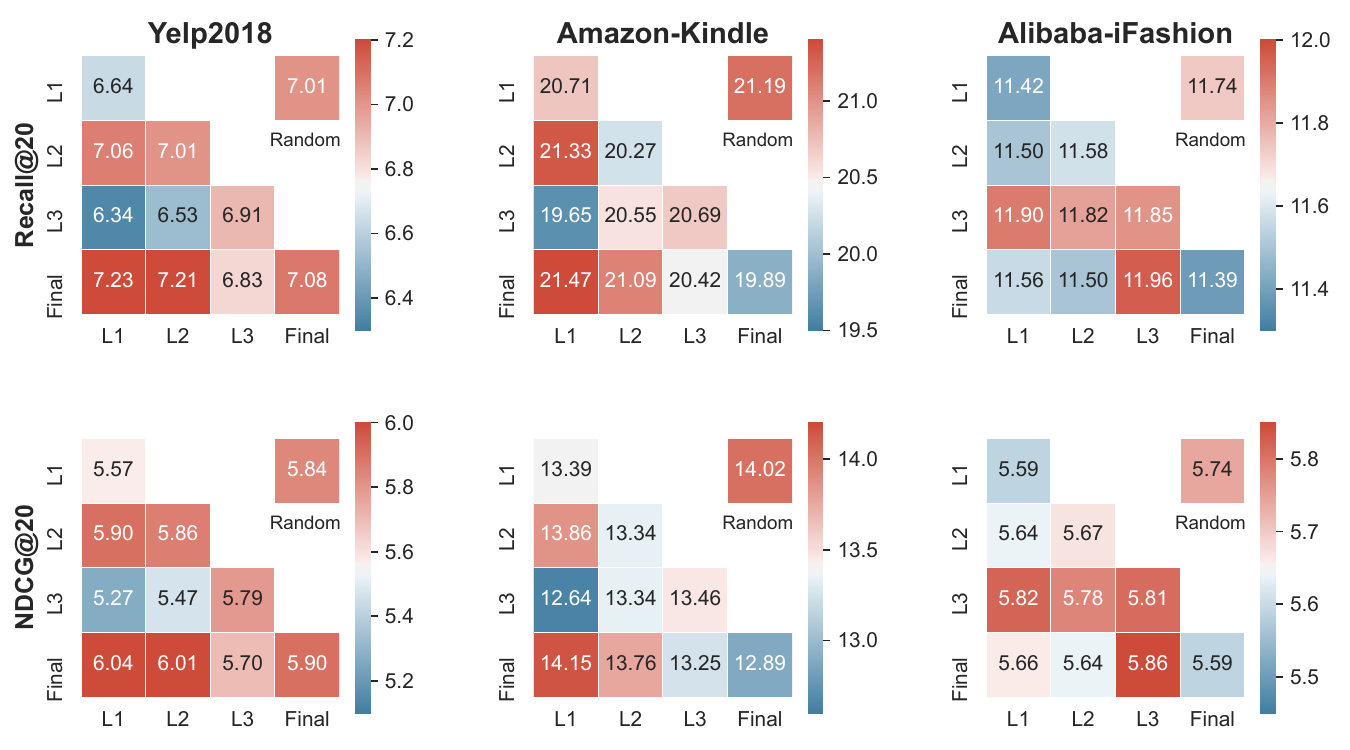}
	\caption{The influence of the layer selection for contrast. }
	\label{fig:layer-contrast}				
\end{figure}

\subsubsection{Layer Selection for Contrast}
In XSimGCL, two layers are chosen to be contrasted. We report the results of different choices in Fig. \ref{fig:layer-contrast} where a 3-layer setting is used. Since these matrix-like heat maps are symmetric, we only display the lower triangular parts. The figures in the diagonal cells represent the results of contrasting the same layer. The optimal layer contrast varies across datasets, but consistently appears to be between the final layer and one of the preceding layers. We analyzed the similarities between representations of different layers and tried to find if $l^{*}$ is related to the similarity but no evidence was found. Fortunately, XSimGCL usually achieves the best performance with a 3-layer setting, which means three attempts are enough. The amount of manual work for tuning $l^{*}$ is therefore greatly reduced. A compromised way without tuning $l^{*}$ is to randomly choose a layer in every mini-batch and contrast its embeddings with the final embeddings. We report the results of this random selection in the upper right of the heatmap. They are acceptable but much lower than the best performance. 

\begin{table}[t]
	\scriptsize
	\caption{Performance comparison of different backbones.}
	\label{Table:nbc}
	\renewcommand\arraystretch{1.0}
	\begin{center}
{
	\begin{tabular}{ccccccc}
		\toprule
		\multirow{2}{*}{\textbf{Method}}&\multicolumn{2}{c}{\textbf{Yelp2018}}& \multicolumn{2}{c}{\textbf{Kindle}} & \multicolumn{2}{c}{\textbf{iFashion}} \cr
		\cmidrule(lr){2-3}\cmidrule(lr){4-5}\cmidrule(lr){6-7}&\textbf{Recall} & \textbf{NDCG}  & \textbf{Recall} & \textbf{NDCG} & \textbf{Recall} & \textbf{NDCG}  \\ \hline
		MF  &\textbf{0.0543}&\textbf{0.0445}& 0.1751& 0.1068 & \textbf{0.0996} & \textbf{0.0468}	\\
		MF + NBC &0.0517&0.0433& \textbf{0.1878}& \textbf{0.1175} & 0.0975 & 0.0453\\
		\hline
		GCN  &0.0556&0.0452& 0.1833& 0.1137 & 0.0952 & 0.0458	\\
		GCN + NBC &\textbf{0.0632}&\textbf{0.0530}& \textbf{0.1989}& \textbf{0.1290} & \textbf{0.1017} & \textbf{0.0486}\\
		\hline
		DNN  &\textbf{0.0522}&\textbf{0.0416}& 0.1243& 0.0705 & 0.0613 & 0.0245	\\
		DNN + NBC &0.0517&0.0413& \textbf{0.1836}& \textbf{0.1213} & \textbf{0.0653} & \textbf{0.0284}\\
		\bottomrule
		\end{tabular}}
	\end{center}
\end{table}

\subsection{Applicability Investigation}
The noise-based CL has been proved effective when combining with LightGCN. We wonder whether this method is applicable to other common backbones such as MF and GCN. Besides, whether uniform noises are the best choice remains unknown. In this part, we examine the applicability of the noise-based augmentation.
\subsubsection{Noised-Based CL on Other Structures}
 We selected three commonly used network structures as backbones: Linear MF, vanilla GCN \cite{kipf2016semi}, and two-tower DNN with two $\mathrm{tanh}$ layers and apply the noised-based CL to them. Since MF cannot adopt cross-layer contrast, we added different uniform noises to the input embeddings for different augmentations. We experimented with various combinations of $\lambda$ and $\epsilon$ on these two structures and report the best results in Table \ref{Table:nbc}, where NBC stands for noise-based CL. Our results demonstrate that NBC can improve the performance of GCN, likely because GCN also has an aggregation mechanism that benefits from contrastive learning. However, NBC cannot consistently improve MF and DNN. On the Amazon-Kindle dataset, obvious improvements are observed whereas on Yelp2018, NBC lowers performance. We will investigate the cause of these inconsistent results in our future work.

 \begin{table}[h]
	\scriptsize
	\caption{Performance comparison of different XSimGCL variants.}
	\label{Table:noise}
	\renewcommand\arraystretch{1.1}
	\begin{center}
{
	\begin{tabular}{ccccccc}
		\toprule
		\multirow{2}{*}{\textbf{Method}}&\multicolumn{2}{c}{\textbf{Yelp2018}}& \multicolumn{2}{c}{\textbf{Kindle}} & \multicolumn{2}{c}{\textbf{iFashion}} \cr
		\cmidrule(lr){2-3}\cmidrule(lr){4-5}\cmidrule(lr){6-7}&\textbf{Recall} & \textbf{NDCG}  & \textbf{Recall} & \textbf{NDCG} & \textbf{Recall} & \textbf{NDCG}  \\ \hline
		LightGCN  &0.0639&0.0525& 0.2057& 0.1315 & 0.1053 & 0.0505	\\
		XSimGCL$_{_a}$ &0.0558&0.0464& 0.1267 & 0.0833 & 0.0158 & 0.0065   \\
		XSimGCL$_{_p}$ &0.0714 & 0.0596& 0.2121 & 0.1398 & 0.1183 & 0.0577   \\
		XSimGCL$_{_g}$ &\underline{0.0722}&\underline{0.0602}& \underline{0.2140} & \underline{0.1410} & \underline{0.1190} & \underline{0.0583}   \\
		XSimGCL &\textbf{0.0723} & \textbf{0.0604}& \textbf{0.2147} & \textbf{0.1415} & \textbf{0.1196} & \textbf{0.0586}   \\ \hline
		w/o CL  &0.0657& 0.0542& 0.1991 & 0.1282 & 0.0973 & 0.0456   \\
		w/o noise &0.0684&0.0573& 0.2048 & 0.1340 & 0.1061 & 0.0515   \\
		w/o both &0.0655 & 0.0540& 0.1990 & 0.1281 & 0.0967 & 0.0450   \\		
		\bottomrule
		\end{tabular}}
		\vspace{-10pt}
	\end{center}
\end{table}

\subsubsection{XSimGCL with Different Noises} 
In this experiment, we test three other types of noises, including adversarial perturbation obtained by following FGSM \cite{goodfellow2014explaining} (denoted by XSimGCL$_a$), positive uniform noises without the sign of learned embeddings (denoted by XSimGCL$_p$), and Gaussian noises (denoted by XSimGCL$_g$). We tried many combinations of $\lambda$ and $\epsilon$ for different types of noises and present the best results in Table \ref{Table:noise}. As observed, the vanilla XSimGCL with signed uniform noises outperforms other variants. Although positive uniform noises and Gaussian noises also bring hefty performance gains compared with LightGCN, adding adversarial noises unexpectedly leads to a large drop of performance. This indicates that only a few particular distributions can generate helpful noises. Additionally, the result that XSimGCL outperforms XSimGCL$_p$ demonstrates the necessity of the sign constraint. In addition to noise types, we also examine whether the added noise would hurt or improve recommendation performance. We present the results of XSimGCL when the contrastive task, the added noise, and both are removed in Table \ref{Table:noise}. The results indicate that without CL, the added noise has little impact on recommendation performance, with only negligible improvements observed. However, when the noise is removed, the contrastive task alone cannot boost the performance to the level of the original XSimGCL, which suggests that both contrastive learning and noise are necessary for a stronger version of XSimGCL. Finally, we want to highlight that \cite{ye2023towards} has validated that noise-based feature perturbation endows SimGCL with the ability to be robust to injected malicious interactions.

\section{Related Work} 
\label{sec:related}
\subsection{GNNs-Based Recommendation Models}
In recent years, graph neural networks (GNNs) \cite{gao2021graph,wu2020graph} have brought the regime of DNNs to an end \cite{chen2020try,wang2018neural,wang2020next}, and become a routine in recommender systems for its extraordinary ability to model the user behavior data \cite{yin2016spatio,yin2015joint}. A great number of recommendation models developed from GNNs have achieved greater than ever performances in different recommendation scenarios \cite{wang2020disentangled,yu2021self,he2020lightgcn,yu2020enhance,wu2019session}. Among numerous variants of GNNs, GCN \cite{kipf2016semi} is the most prevalent one and drives many state-of-the-art graph neural recommendation models such as NGCF \cite{wang2019neural}, LightGCN \cite{he2020lightgcn}, LR-GCCF \cite{chen2020revisiting} and LCF \cite{yu2020graph}. Despite varying implementation details, all these GCN-based models share a common scheme which is to aggregate information from the neighborhood in the user-item graph layer by layer \cite{wu2020graph}. Benefitting from its simple structure, LightGCN becomes one of the most popular GCN-based recommendation models. It follows SGC \cite{wu2019simplifying} to remove the redundant operations in the vanilla GCN including transformation matrices and nonlinear activation functions. This design is proved efficient and effective for recommendation where only the user-item interactions are provided. It also inspires a lot of CL-based recommendation models such as SGL \cite{wujc2021self}, NCL \cite{lin2022improving} and SimGCL \cite{yu2022graph}.

\subsection{Contrastive Learning for Recommendation}
Contrastive learning \cite{jaiswal2021survey,liu2020self} recently has drawn considerable attention in many fields due to its ability to deal with massive unlabeled data \cite{gao2021simcse,chen2020simple,you2020graph}. As CL usually works in a self-supervised manner \cite{yu2022survey}, it is inherently a sliver bullet to the data sparsity issue \cite{yu2018adaptive} in recommender systems. Inspired by the success of CL in other fields, the community also has started to integrate CL into recommendation \cite{zhou2020s,wujc2021self,xia2020self,yu2021socially,yu2021self,ma2020disentangled,zhou2021contrastive}. To the best of our knowledge, S$^{3}$-Rec \cite{zhou2020s} is the first work that combines CL with sequential recommendation. It first randomly masks part of attributes and items to create sequence augmentations, and then pre-trains the Transformer \cite{vaswani2017attention} by encouraging the consistency between different augmentations. The similar idea is also found in a concurrent work CL4SRec \cite{xie2022contrastive}, where more augmentation approaches including item reordering and cropping are used. Besides, S$^{2}$-DHCN \cite{xia2020self} and ICL \cite{lin2022improving} adopt advanced augmentation strategies by re-organizing/clustering the sequential data for more effective self-supervised signals. Qiu \textit{et al.} proposed DuoRec \cite{qiu2022contrastive} which adopts a model-level augmentation by conducting dropout on the encoder. Xia \textit{et al.} \cite{xia2022device,xia2023efficient} integrated CL into a self-supervised knowledge distillation framework to transfer more knowledge from the server-side large recommendation model to resource-constrained on-device models to enhance next-item recommendation.   
In the same period, CL was also introduced to different graph-based recommendation scenarios. S$^{2}$-MHCN \cite{yu2021self} and SMIN \cite{long2021social} integrate CL into social recommendation. HHGR \cite{zhang2021double} proposes a double-scale augmentation approach for group recommendation and develops a finer-grained contrastive objective for users and groups. CCDR \cite{xie2022kdd} has explored the use of CL in cross-domain and bundle recommendation. Yao \textit{et al.} \cite{yao2021self} proposed a feature dropout-based two-tower architecture for large-scale item recommendation. NCL \cite{lin2022improving} designs a prototypical contrastive objective to capture the correlations between a user/item and its context. SEPT \cite{yu2021socially} and COTREC \cite{xia2021self} further propose to mine multiple positive samples with semi-supervised learning on the perturbed graph for social/session-based recommendation. The most widely used model is SGL \cite{wujc2021self} which replies edge/node dropout to augment the graph data. Although these methods have demonstrated their effectiveness, they pay little attention to why CL can enhance recommendation.

\section{Conclusion}\label{sec:conclusion}
In this paper, we revisit the graph CL in recommendation and investigate how it enhances graph recommendation models. The findings are surprising that the InfoNCE loss is the decisive factor which accounts for most of the performance gains, whilst the elaborate graph augmentations only play a secondary role. Optimizing the InfoNCE loss leads to a more even representation distribution, which helps to promote the long-tail items in the scenario of recommendation. In light of this, we propose a simple yet effective noise-based augmentation approach, which can smoothly adjust the uniformity of the representation distribution through CL. An extremely simple model XSimGCL is also put forward, which brings an ultralight architecture for CL-based recommendation. The extensive experiments on four large and highly sparse datasets demonstrate that XSimGCL is an ideal alternative of its graph augmentation-based counterparts.\par


	\ifCLASSOPTIONcaptionsoff
	\newpage
	\fi
	\bibliographystyle{IEEEtran}
	\bibliography{refs}
	
	\begin{IEEEbiography}[{\includegraphics[width=1in,height=1.25in,clip,keepaspectratio]{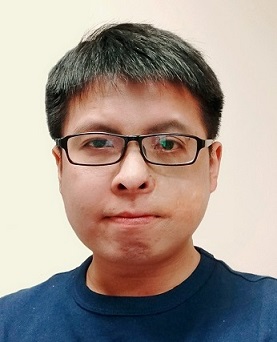}}]{Junliang Yu}
		completed his B.S. and M.S degrees at Chongqing University, and PhD degree at The University of Queensland. Currently, he is a postdoctoral research fellow at the School of Information Technology and Electrical Engineering, the University of Queensland. His research interests include recommender systems, tiny machine learning, and self-supervised learning.
	\end{IEEEbiography}
		
	\begin{IEEEbiography}[{\includegraphics[width=1in,height=1.25in,clip,keepaspectratio]{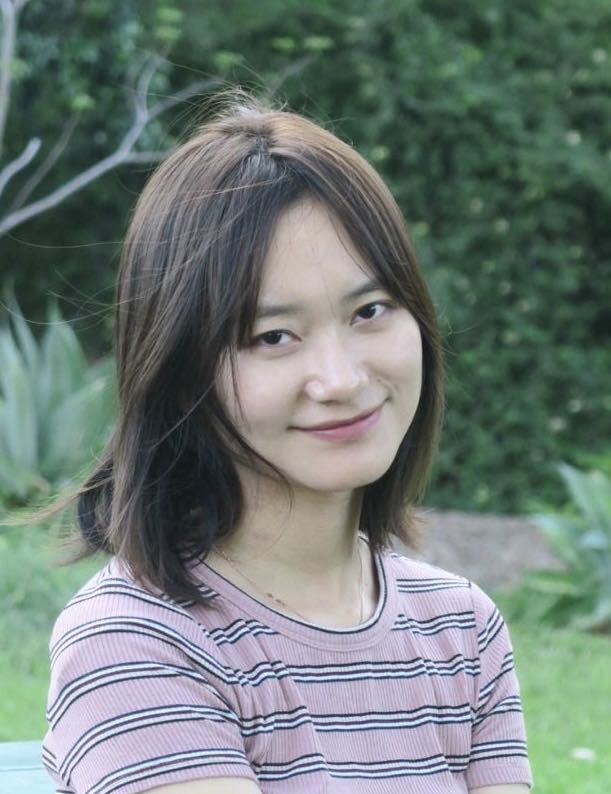}}]{Xin Xia}
		received her B.S. degree in Software Engineering from Jilin University, China. Currently, she is a final-year Ph.D. candidate at the School of Information Technology and Electrical Engineering, the University of Queensland. Her research interests include on-device machine learning, sequence modeling, and self-supervised learning.
	\end{IEEEbiography}

	\begin{IEEEbiography}[{\includegraphics[width=1in,height=1.25in,clip,keepaspectratio]{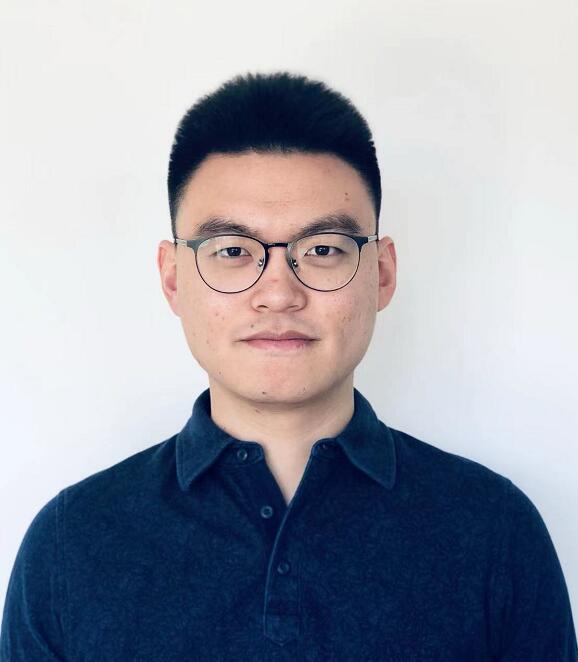}}]{Tong Chen}
		received his PhD degree in computer science from The University of Queensland in 2020. He is currently a Lecturer with the Data Science research group, School of Information Technology and Electrical Engineering, The University of Queensland. His research interests include data mining, recommender systems, user behavior modelling and predictive analytics.
	\end{IEEEbiography}

	\begin{IEEEbiography}[{\includegraphics[width=1in,height=1.25in,clip,keepaspectratio]{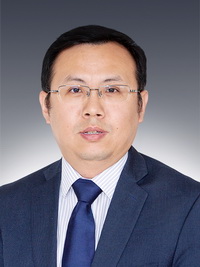}}]{Lizhen Cui}
		is a full professor with Shandong University. He is appointed dean and deputy party secretary for School of Software, co-director of Joint SDU-NTU Centre for Artificial Intelligence Research(C-FAIR), director of the Research Center of Software and Data Engineering, Shandong University. His main interests include big data intelligence theory, data mining, wisdom science, and medical health big data AI applications.
	\end{IEEEbiography}

	\begin{IEEEbiography}[{\includegraphics[width=1in,height=1.25in,clip,keepaspectratio]{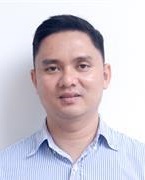}}]{Nguyen Quoc Viet Hung}
		is a senior lecturer and an ARC DECRA Fellow in Griffith University. He earned his Master and PhD degrees from EPFL in 2010 and 2014 respectively. His research focuses on Data Integration, Data Quality, Information Retrieval, Trust Management, Recommender Systems, Machine Learning and Big Data Visualization, with special emphasis on web data, social data and sensor data. 
	\end{IEEEbiography}

	\begin{IEEEbiography}[{\includegraphics[width=1in,height=1.25in,clip,keepaspectratio]{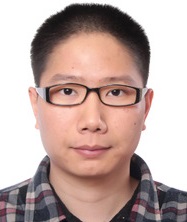}}]{Hongzhi Yin}
		is an associate professor and ARC Future Fellow at the University of Queensland. He received his Ph.D. degree from Peking University, in 2014. His research interests include recommendation system, deep learning, social media mining, and federated learning. He is currently serving as Associate Editor/Guest Editor/Editorial Board for ACM Transactions on Information Systems (TOIS), ACM Transactions on Intelligent Systems and Technology (TIST), etc.
   \end{IEEEbiography}
	%
	%
	%
	
	
	

\end{document}